\documentclass[11pt,letter]{article}
\pdfoutput=1 %% activate this line when submit %%

\usepackage[dvipsnames,usenames,svgnames]{xcolor}
\usepackage{jheppub}
\usepackage{amssymb,amsmath,amsthm,graphicx}
\usepackage{mathrsfs}
\usepackage{graphics, color}
\usepackage{subfigure}
\usepackage{latexsym}
\usepackage{bm}
\usepackage[hang,flushmargin]{footmisc}
\usepackage{epsfig}
\usepackage{braket}
\usepackage{textcomp}
\hypersetup{
 pdfauthor={David Grabovsky, Maciej Kolanowski},citecolor=PineGreen,linkcolor=Maroon,urlcolor=RoyalBlue}
\usepackage[utf8]{inputenc}
\allowdisplaybreaks[1]

\hypersetup{
 pdfauthor={David Grabovsky},
 urlcolor=PineGreen, citecolor=Maroon, linkcolor=RoyalBlue}

\newcommand{\f}[2]{\frac{#1}{#2}}

\def \dd {\mathrm{d}}

\begin{document}
%---Title and Abstract -----------------------------------------------------------
\title{Spin-Refined Partition Functions and $\mathcal{CRT}$ Black Holes}
\author{David~Grabovsky and Maciej~Kolanowski}

\affiliation{Department of Physics, University of California, Santa Barbara, CA 93106, U.S.A.}
\emailAdd{davidgrabovsky@ucsb.edu}
\emailAdd{mkolanowski@ucsb.edu}

\newcommand{\blue}{\color{blue}}
\newcommand{\red}{\color{red}}

\abstract{
We investigate spin-refined partition functions in AdS/CFT using Euclidean gravitational path integrals. We construct phase diagrams for $Z_X = \text{Tr} \big( e^{-\beta H} X \big)$ in various dimensions and for different choices of discrete isometry $X$, discovering rich structures at finite temperature. When $X$ is a reflection, $Z_X$ counts the difference between the number of even- and odd-spin microstates. The high-temperature regime is universally dominated by $\mathcal{CRT}$-twisted black holes in any dimension, and in odd spacetime dimensions we examine whether complex rotating black hole solutions can contribute to spin-refined observables or potentially dominate at finite temperature. We also analyze the microcanonical ensemble. There the leading contribution almost always comes from rotating black holes, showing that the two ensembles are not necessarily equivalent. 
}

\maketitle

\newpage
%%%%%%%%%%%%%%%%%%%%%
\section{Introduction} \label{sec:Introduction}

In recent years there has been a surge of activity in trying to understand black hole microstates using gravitational path integrals \cite{Gibbons:1976ue, Cabo-Bizet:2018ehj, Iliesiu:2022kny, Iliesiu:2022onk, LopesCardoso:2022hvc, H:2023qko,Sen:2023dps,Anupam:2023yns}, especially in AdS/CFT \cite{Maldacena:1997re, Witten:1998zw}. The main interest lies in computing  the thermal partition function $Z = \textrm{Tr} \big( e^{-\beta H} \big)$ and the density of states, but these are not the only probes of statistics that one can employ. The most famous alternative is the Witten index $\textrm{Tr} \big( e^{-\beta H} (-1)^{\mathsf{F}} \big)$ \cite{Witten:1982df}, along with its generalizations (see for example \cite{Sen:2012hv}). The index counts the difference between the number of bosonic and fermionic states in the theory by tracing over fermions with the opposite sign. An even more general quantity was considered in \cite{Benjamin:2019stq}, where the twisted partition function $\textrm{Tr} \big( e^{-\beta H + \alpha J} \big)$
was studied in 2d conformal field theories (CFTs), with special attention given to the case $\alpha \in 2\pi i \mathbb{Q}$. Observables of this form, which we will call \textit{spin-refined partition functions}, have the potential to give more fine-grained insight into the microstate structure of gravitational theories. For similar results in the presence of global and finite-group symmetries, see \cite{Pal:2020wwd,Harlow:2021trr}.

To be more concrete, our starting point is thermal partition function in $D$ spacetime dimensions, with a discrete rotation inserted: we consider
\begin{equation}
\label{starting_point}
Z_{\vec{\alpha}}(\beta) = \textrm{Tr} \left( e^{-\beta H + \vec{\alpha} \cdot \vec{J}} \right),
\end{equation}
where $\vec{J} = (J_1, ..., J_n)$ is the angular momentum, taking values in the Cartan subalgebra of the rotation algebra $\mathfrak{so}(D-2)$, and $\vec{\alpha}$ is a set of fixed numbers. Most often we will assume that all of them are rational multiples of $2\pi i$. To get a better feeling for the quantities we will be studying, let us start by providing a microscopic interpretation of
\begin{equation}
\label{thermal_3d}
    Z_R(\beta) = \textrm{Tr} \left( e^{-\beta H + \pi i J} \right)
\end{equation}
in 2d CFTs. If the theory living on the boundary has reflection symmetry, then for every eigenvector $\ket{E,j}$ of the Hamiltonian $H$ and angular momentum $J$, there exists also an eigenvector $\ket{E,-j}$. If $j \in \frac{1}{2} + \mathbb{Z}$, the contributions from $j$ and $-j$ differ by a phase that cancels them out exactly. Thus, $Z_R$ is a purely bosonic quantity.\footnote{Note that this does not mean that the presence of fermions in the theory or lack thereof is meaningless. Rather, only bosonic states contribute, but these states may still be built out of fermions.} For bosons with even $j$, $e^{\pi i J}$  is positive, and for bosons with odd $j$ it is negative. Thus we may rewrite $Z_R$ as 
\begin{equation}
\label{density_diff}
    Z_{R}(\beta) = \big(\textrm{Tr}_{j \textrm{\ even}} - \textrm{Tr}_{j \textrm{\ odd}} \big) e^{-\beta H}.
\end{equation}

Notice that on $S^1$, the antipodal map is the same as the rotation $\phi \mapsto \phi + \pi$. More generally, it is a property of odd-dimensional spheres that the antipodal map is just a rotation by $\pi$. Thus, in this case we have $Z_R(\beta) = \textrm{Tr} \big( e^{-\beta H} R \big)$, where $R$ is the reflection operator.\footnote{We use $R$ to denote an insertion that reflects of all of the coordinates. Thus $R = \cal{RT}$, where $\cal{R}$ and $\cal{T}$ are the discrete symmetries corresponding to spatial reflection and (Euclidean) time reversal.} We will in fact take this as the definition of $Z_R$ in any dimension. We may still interpret it as \eqref{density_diff}, where now the labels ``even'' and ``odd'' refer to the properties of states under $R$. If the boundary theory has $R$ as a symmetry, then $Z_R$ will be a bosonic quantity in any dimension.

It was recently argued that the gravitational path integral should include Lorentzian geometries that are not time-orientable \cite{Harlow:2023hjb}. This is a logical consequence of the lack of global symmetries in quantum gravity. It is a well-known theorem that $\mathcal{CRT}$\footnote{$\mathcal{CRT}$ stands for $\cal{C}$harge conjugation, spatial $\cal{R}$eflection, and $\cal{T}$ime reversal. It is most naturally defined in the Euclidean signature as a rotation by $\pi$. In the older literature, this operation is often called $CPT$.} must be a symmetry of any relativistic quantum field theory \cite{Streater:1989vi}. Nevertheless, in any consistent theory of quantum gravity we must either gauge it or break it: \cite{Harlow:2023hjb} chose the former alternative. As advertised, this leads to the natural possibility that the gravitational path integral could receive important contributions from non-time-orientable spacetimes. It was further argued in \cite{Harlow:2023hjb} that from a holographic point of view, such saddles should give the dominant contribution to \eqref{thermal_3d} in 3d gravity. We will study this question also in higher dimensions.

Let us pause to emphasize that in our usage, the operator $R$ changes the signs of all the coordinates. This may be confusing, because in Euclidean signature the parity operator flips the signs of all spatial coordinates, while the reflection operator flips only one of them. But parity and reflection are actually the same in odd bulk spacetime dimensions, up to a rotation. (For example, in $\text{AdS}_3/\text{CFT}_2$ they are identical.) Our focus in this work will be on examples in odd spacetime dimensions, so we will not be very careful with the distinction.\footnote{The reflection of only one spatial coordinate has determinant $-1$ in any dimension, so Euclidean saddles for that insertion can be non-orientable. Such objects have been studied in $\text{AdS}_3/\text{CFT}_2$ \cite{Maloney:2016gsg,Wei:2024zez}, but these are not the saddles we are looking for. We thank Zixia Wei for clarifying comments on this point.} In practice, we take the definition of $R$ to be such that it implements (\ref{density_diff}).

\subsection{Review of the $\mathcal{CRT}$-twisted black hole}

The authors of \cite{Harlow:2023hjb} provided an example of a universal saddle contributing to $Z_R$ that we will now review. Let us start with the non-rotating BTZ black hole with inverse temperature $\beta$. In Kruskal-Szekeres coordinates $(X^{\pm}, \phi)$, the metric reads
\begin{equation}
    ds^2 = -\frac{4 dX^+ dX^-}{(1+X^+ X^-)^2} + \left(
    \frac{2\pi}{\beta} \frac{1-X^+ X^-}{1+X^+ X^-}
    \right)^2 d\phi^2.
\end{equation}
The Euclidean action of this solution is
\begin{equation}
\label{eqn:BTZ-action}
    I_{\textrm{BTZ}}(\beta) = - \frac{\pi^2 c}{3\beta},
\end{equation}
where $c$ is the central charge of the dual CFT. Let us divide this saddle by the following $\mathbb{Z}_2$:
\begin{equation}
    (X^+, X^-, \phi) \mapsto (-X^+, -X^-, \phi + \pi).
\end{equation}
This $\mathbb{Z}_2$ action has no fixed points and thus the resulting manifold, called the $\mathcal{CRT}$-twisted black hole, is smooth. As a quotient of BTZ, it still satisfies the Einstein equations. Its Euclidean action is simply half of (\ref{eqn:BTZ-action}),
\begin{equation}
    I_{\mathcal{CRT}}(\beta) = - \frac{\pi^2 c}{6\beta}.
\end{equation}

Having established that the $\mathcal{CRT}$-twisted black hole is a saddle for the gravitational path integral, we should also establish what boundary conditions it satisfies. The time circle gets cut in half, so its inverse temperature is $\frac{\beta}{2}$. Moreover, the identification introduces a twist in a spatial circle. It was argued in \cite{Harlow:2023hjb} that this saddle contributes to $\textrm{Tr} \big( e^{-\frac{\beta}{2} H + \pi i J} \big)$.\footnote{It may be worth emphasizing that this is not an index, since $(-1)^{\mathsf{F}} = e^{2\pi i J}$.} Notice that this corresponds to the modular parameter $\tau = \frac{1}{2} + i \frac{\beta}{4\pi}$. The partition function at this value of the modular parameter can be rewritten using modular invariance as
\begin{equation}
    Z(\tau) = Z\left(\frac{-\tau}{2 \tau -1}\right) = Z\left(-\frac{1}{2} + i \frac{\pi}{\beta}\right).
\end{equation}
Thus the high-temperature regime can be mapped to the low-temperature regime, where the partition function is dominated by the thermal AdS$_3$. One can easily see that $I_{\mathcal{CRT}}$ reproduces that result quantitatively at small $\beta$. As was pointed in \cite{Chen:2023mbc}, that saddle is in fact a well-known solution, namely one of $SL(2,\mathbb{Z})$ black holes.

We may generalize $\mathcal{CRT}$-twisted black hole to any dimension. Indeed, let us start with the Schwarzschild AdS (SAdS) black hole and quotient it by
\begin{equation}
\label{eqn:CRT-quotient}
    (X^+, X^-, y^a) \mapsto (-X^+, -X^-, -y^a),
\end{equation}
where $(X^+, X^-, y^a)$ are Kruskal-Szekeres coordinates and $-y^a$ represents the antipodal map on a sphere. Since the action of this $\mathbb{Z}_2$ is free, the resulting manifold is always smooth. It is easy to calculate the Euclidean action of the quotient:
\begin{equation}
    I_{\mathcal{CRT}}\left(\frac{\beta}{2}\right) = \frac{1}{2} I_{\textrm{SAdS}}(\beta). \label{I_CRT}
\end{equation}
Since at large temperatures in $D$ dimensions we have $I_{\textrm{SAdS}}(\beta) \sim \beta^{2-D}$, it follows that
\begin{equation}
\label{I_CRT-high-temp}
    I_{\mathcal{CRT}}(\beta)= \frac{1}{2^{D-1}} I_{\textrm{SAdS}}(\beta) + O \left( \beta^{3-D} \right).
\end{equation}
Thus, the $\mathcal{CRT}$ saddle exactly reproduces high-temperature behavior of the free energy of even-spin minus odd-spin operators that was obtained in the thermal effective theory of \cite{Benjamin:2024kdg}.

A few observations regarding that problem are in place. One may be worried that a large split between even and odd $j$ could lead to problems with the semiclassical limit of the bulk theory. After all, we do not expect that the scattering process of a photon on the Kerr background should depend significantly on whether the spin of the black hole is $j$ or $j + \hbar$. However, while $Z_R(\beta)$ may be large, the relative change in the density of states $Z_R(\beta)/Z(\beta)$ remains exponentially small as $G_{\text{N}} \to 0$. Thus, the semiclassical physics should be agnostic to the even-odd split.

Another natural question is what this split really looks like in the density of states. Are there systematically more even than odd spin states?\footnote{To be more precise, so far we have only established that there is only a large difference between even and odd spin states. The sign of that difference would be fixed by one-loop determinants and thus may be matter-content dependent. A priori, the phase could be also temperature-dependent, but this is rather unlikely.} Or maybe the difference oscillates rapidly and only after averaging over a sufficiently wide energy window, we obtain a definite split? We will answer these questions in the rest of this paper.

It was already pointed out in \cite{Chen:2023mbc} that the three-dimensional $\mathcal{CRT}$-twisted black hole is simply one of many $SL(2,\mathbb{Z})$ black holes present in $\text{AdS}_3$. One may be worried that the lack of time-orientability in the Lorentzian signature may translate to a lack of orientability in the Euclidean one. Interestingly, since all of the time-orientability issues are hidden behind the horizon, a Euclidean path integral would not be able to detect any causality-related problems with that saddle.

The rest of this paper is organised as follows. In Sec. \ref{sec_3d}, we discuss spin-refined partition functions in AdS$_3$/CFT$_2$. We start with a general review of the thermodynamics of systems with angular momenta, valid in arbitrary dimension, and then work out in detail the example of $SL(2,\mathbb{Z})$ black holes. From there, we will construct a few phase diagrams for spin-refined partition functions. In Sec. \ref{5d}, we perform similar analysis for AdS$_5$/CFT$_4$. We finish with future directions and some additional observations in Sec. \ref{sec:Discussion}.

When this work was nearing completion, we learned of the work \cite{Benjamin:2024kdg}, which starts with \eqref{starting_point} and evaluates it at small $\beta$ using a thermal effective field theory. Our work, though closely related, is more gravitational in nature and takes a complementary approach.

%%%%%%%%%%%%%%%%%%%%%
\section{Rotating black holes in 3d gravity} \label{sec_3d}
\subsection{Thermodynamics of rotating systems}
One traditionally starts discussions of thermodynamics with the entropy $S$ as a function of extensive parameters\footnote{Although these parameters are not necessarily extensive in black hole thermodynamics, the same ideas apply.}---in our case, the energy $E$ and the angular momentum $J$.\footnote{In general spacetime dimension $D$, angular momentum takes values in the Cartan subalgebra of the rotation algebra $\mathfrak{so}(D-2)$, and black hole thermodynamics is modified accordingly.} The first law of black hole thermodynamics then reads
\begin{equation}
    \dd S = \beta\, \dd E - \beta \Omega\, \dd J,
\end{equation}
where $\beta$ is the inverse temperature and $\Omega$ is the angular velocity.
Usually one then goes to define the free energy as
\begin{equation}
    F = E - \beta^{-1}S,
\end{equation}
written as a function of $\beta$ and $J$. This is not what we want to do. We want to fix both $\beta$ and $\Omega$, and to this end we define
\begin{equation}
    F_{\text{rot}} = E - \beta^{-1}S - \Omega J.
\end{equation}
It is easy to check that
\begin{equation}
    \dd F_{\text{rot}} = - S\, \dd T -  J\, \dd \Omega.
\end{equation}
Moreover, $F_{\text{rot}}$ can be defined statistically, i.e. in terms of the microstates, as
\begin{equation}
    e^{-\beta F_{\text{rot}}} := \textrm{Tr}
\left( e^{-\beta H+ \beta\Omega J} \right).
\end{equation}
We will derive all of the identities we need in terms of real $\Omega$, and then analytically continue it to the imaginary axis. This can be implemented in a path integral by imposing following boundary conditions on the geometry at infinity:
\begin{equation}
    (t_E, \phi) \sim (t_E + \beta, \phi + i \beta \Omega) \sim (t_E, \phi+2\pi),
\end{equation}
where $t_E$ is the Euclidean time and $J$ generates the rotation along the $\phi$ direction.

We are also interested at the microcanonical formalism, where one defines $S_{\text{rot}}$ by
\begin{equation}
    e^{S_{\text{rot}}} := \textrm{Tr}_E \left( e^{\beta \Omega J} \right).
\end{equation}
The trace is taken over states with fixed energy $E$, such that for each state $\beta \Omega$ is fixed to a prescribed value.\footnote{The notation $\mathrm{Tr}_E$ means that we trace over states in a small window of energies $(E, E + \delta E)$. This typically gives microcanonical quantities which agree with Legendre transforms of appropriate canonical quantities. However, oscillatory quantities like $e^{\pi i J}$ are not positive-definite, so changing the window slightly can alter the results. However, with a sufficiently wide window, there is a natural sense in which such quantities oscillate around a central value. This is the perspective taken in \cite{Chen:2023mbc}, which also tried to quantify the window-dependence of oscillatory quantities by studying wormhole contributions to their connected correlator.}
We have, then,
\begin{equation}
    S_{\text{rot}} = S + \beta \Omega J,
\end{equation}
and one can easily check that 
\begin{equation}
    \dd S_{\text{rot}} = \beta\, \dd E + J\, \dd \left(\beta \Omega \right).
\end{equation}
Therefore $S_{\text{rot}}$ is in a natural way a function of $E$ and $\beta \Omega =: \alpha$, which is consistent with the fact that in the trace expression we keep that combination fixed.

Let us finish this introduction by applying the formalism to the $\mathcal{CRT}$ black hole. Using \eqref{I_CRT} and standard thermodynamical relations, we get
\begin{equation}
    S_{CRT}(\beta) = \frac{1}{2} S_{\textrm{SAdS}}(2\beta).
\end{equation}
Since the $\mathcal{CRT}$ black hole is a quotient of a static black hole, it does not rotate. Thus, it follows that $S_{\mathcal{CRT}} =S_{\mathcal{CRT}, \textrm{rot}}$ and also
\begin{equation}
    E_{\mathcal{CRT}}(\beta) = E_{\textrm{SAdS}}(2\beta).
\end{equation}
By inverting the relationship between $E$ and $\beta$, we finally arrive at the conclusion
\begin{equation}
\label{eqn:CRT-entropy}
    S_{\mathcal{CRT}}(E) = \frac{1}{2} S_{\textrm{SAdS}}(E).
\end{equation}
It follows in particular that in any dimension, the $\mathcal{CRT}$ black hole contribution to $\textrm{Tr}_E R$ is equal to $\sqrt{\textrm{Tr}_E 1}$. However, as we will see below, this is far from being the dominant contribution, at least at large energies.

\subsection{An example: BTZ black holes}
In the ensemble of fixed temperature and angular velocity, which we will call the canonical ensemble from now on, the free energy of the rotating BTZ black hole reads
\begin{equation}
    F_{\text{rot}} = \frac{\pi^2 c}{3\beta^2 (\Omega^2 -1)}.
\end{equation}
From this we may easily calculate the entropy\footnote{One may wonder why we bother deriving (for example) the entropy, since this is proportional to the horizon's area and thus easily obtained from the geometry. The reason is that we treat it as a warm-up before the case of a general $SL(2,\mathbb{Z})$ black hole, where these expressions may be less familiar.}
\begin{equation}
    S = -\left(\frac{\partial F_{\text{rot}}}{\partial T} \right)_{\Omega} = \frac{2 \pi^2 c}{3\beta (1 - \Omega^2)},
\end{equation}
the angular momentum
\begin{equation}
    J = - \left(\frac{\partial F_{\text{rot}}}{\partial \Omega} \right)_\beta = \frac{2 \pi^2 \Omega c}{3\beta^2 (1 - \Omega^2)^2},
\end{equation}
and the energy
\begin{equation}
    E = F + T S + \Omega J = \frac{\pi^2 (\alpha^2 + \beta^2) c}{3 (\alpha^2 - \beta^2)^2},
\end{equation}
where we have substituted $\alpha := \Omega \beta$. This equation can be solved for $\beta(E, \alpha)$ (there are in fact four solutions that we may denote by $\beta_{\pm \pm}$) and then plugged into
\begin{equation}
    S_{\text{rot}} = S + \alpha J
\end{equation}
treated as a function of $E$ and $\alpha$. The analytical result is not very illuminating. What actually matters to us is the behavior of $S_{\text{rot}}$ as $E \to \infty$. We find that for $\alpha^2 E \gg c$, we have
\begin{equation}
\label{eqn:Srot-BTZ}
    S_{\text{rot}, \pm \pm}(E, \alpha) = \pm \alpha E \pm \frac{1}{\sqrt{2}} S_0(E) + O(1),
\end{equation}
where $S_0(E) = S_{rot, ++}(E,0)= 2 \pi \sqrt{\frac{c}{3}(E - \frac{c}{12})}$ is the entropy without any rotation \cite{Cardy:1986ie}.

Let $\alpha = i\tilde{\alpha}$, so that $\tilde{\alpha} \in \mathbb{R}$. The quantity we are actually interested in is $\textrm{Tr}_E \big( e^{i \tilde\alpha J}\big)$, and this is determined by $e^{S_{\text{rot}}}$ for the branch of (\ref{eqn:Srot-BTZ}) with the largest real part. The first term in $S_{\text{rot}, \pm \pm}$ is purely imaginary and describes oscillations in the density of states, while the second term is the leading real part. The solutions $S_{\text{rot}, \pm, +}$ are dominant, while $S_{\text{rot}, \pm -}$ are exponentially suppressed. Thus, at large energies we may write
\begin{equation} \label{micro_BTZ}
    \textrm{Tr}_{E} \left( e^{i \tilde{\alpha} J} \right) = 2 \cos(\tilde\alpha E) \exp\left(\frac{1}{\sqrt{2}} S_0(E)\right).
\end{equation}
This is not yet the full answer, since we need to consider all of the $SL(2, \mathbb{Z})$ black holes.

First, however, note that $\tilde{\alpha}$ is defined only mod $2 \pi$.\footnote{or $4\pi$, if the theory includes fermions. We will discuss the inclusion of fermions below.} It follows (for example, from modular invariance) that we should include all possible discrete values of $\tilde{\alpha}$. If $\tilde{\alpha}$ happens to be $0$ (meaning the angular velocity is zero), then saddles with $\tilde{\alpha} + 2\pi n$  $(n \neq 0)$ will be subleading. However, if $\tilde{\alpha} \notin 2 \pi \mathbb{Z}$, then all possible values of $\tilde{\alpha}$ contribute at the same order: the difference between them is $O\big(\frac{1}{\sqrt{E}}\big)$.
%%%%%%%%%%%%%%%%%%%%%
\subsection{The $SL(2, \mathbb{Z})$ black holes}

We are now ready to discuss the thermodynamics properties of the $SL(2, \mathbb{Z})$ black holes. For a classic reference (which includes also the one-loop determinants for these saddles), see \cite{Maloney:2007ud}. The $SL(2, \mathbb{Z})$ black holes $M_{C,D}$ are parametrized by the $C$ and $D$ parameters of the modular transformation\footnote{We use capital letters here to avoid confusion with the central charge $c$.}
\begin{equation}
    \tau \mapsto \frac{A\tau + B}{C\tau+D},
\end{equation}
where $A,B,C,D \in \mathbb{Z}$ and $AD-BC =1$. Since the modular group is $SL(2,\mathbb{Z})/\mathbb{Z}_2$, we can restrict to $C \ge 0$. For example, the BTZ black hole corresponds to $M_{1,0}$, while thermal AdS is $M_{0,1}$. The semiclassical contribution to the  partition function associated to the $M_{C,D}$ black hole with modular parameter $\tau = \frac{\tilde{\alpha}}{2\pi} + \frac{i\beta}{2 \pi}$ is given by
\begin{equation}
    Z_{C,D}(\tau) = Z_{\text{AdS}} \left( 
\frac{A\tau + B}{C\tau+D}
    \right) = \exp \left(
\frac{\pi c}{6} \Im \left(\frac{A\tau + B}{C\tau+D}
    \right)\right) = \exp \left(
    \frac{\pi^2 c}{3} \frac{\beta}{(\tilde\alpha C + 2\pi D)^2 + C^2 \beta^2}
    \right).
\end{equation}

It follows that the free energy reads
\begin{equation}
    F_{\text{rot}} = -\frac{\pi^2 c}{3} \frac{1}{(\tilde\alpha C + 2\pi D)^2 + C^2 \beta^2}.
\end{equation}
Notice that when $\tilde\alpha \in \pi\mathbb{Q}$, there is a special modular transformation such that $\tilde\alpha C + 2 \pi D = 0$. We will come back to this shortly. To keep the `real' notation from the previous section, write $i \tilde{\alpha} = \alpha = \beta \Omega$. We may use standard thermodynamical relations to get
\begin{subequations}
    \begin{equation}
       J= -\left(\frac{\partial F_{\text{rot}}}{\partial\Omega} \right)_\beta = \frac{\pi^2 c}{3} \frac{2 i \beta C (2 \pi D-i C \beta \Omega )}{\big(\beta ^2 C^2+(2 \pi  D-i C \beta \Omega )^2\big)^2},
    \end{equation}
    \begin{equation}
        S = -\left(\frac{\partial F_{\text{rot}}}{\partial T} \right)_\Omega =- \frac{\pi^2 c}{3} \frac{2 \beta ^2 C \left(\beta  C \left(\Omega ^2-1\right)+2 \pi i D \Omega \right)}{ \left(\beta ^2 C^2 \left(\Omega ^2-1\right)+4 \pi i  \beta  CD \Omega -4 \pi ^2 D^2\right)^2},
    \end{equation} and
    \begin{equation}
        E = F_{\text{rot}} + TS + \Omega J = \frac{\pi^2 c}{3} \frac{C^2 \left(\alpha ^2+\beta ^2\right)+4 \pi i  \alpha  CD-4 \pi ^2 D^2}{\left(C^2 \left(\beta ^2-\alpha ^2\right)-4 \pi i \alpha  CD+4 \pi ^2 D^2\right)^2}.
    \end{equation}
    \end{subequations}
As before, the last equation (where we have put $\alpha = \beta\Omega$) can be solved for $\beta(E, \alpha)$ (there are again four solutions $\beta_{\pm \pm}$) and then plugged into $S_{\text{rot}} = S + \alpha J$, treated as a function of $E$ and $\alpha$. The behavior of $S_{\text{rot}}$ as $E \to \infty$, for generic values of $\alpha, C,D$ is
\begin{equation}
\label{eqn:Srot-SL2}
    S_{\text{rot},\pm\pm}(E,\alpha) = \pm i \left(\frac{2D \pi - i C \alpha}{C}\right) E \pm \frac{1}{\sqrt{2} C} S_0(E) + O(1).
\end{equation}

Setting $\alpha = i\tilde{\alpha}$, we see that once again the first term above is purely imaginary and describes oscillations. The second term is the leading real part, and as before $S_{\text{rot}, \pm +}$ is dominant while $S_{\text{rot}, \pm -}$ is suppressed. Notice, however, that for $C>1$, $ S_{\text{rot},\pm\pm}(E,\alpha)$ is always subleading with respect to $C=1$ (which corresponds to the BTZ black hole). Thus, at large energies the microcanonical ensemble is always dominated by the BTZ black hole.

Let us now consider the fine-tuned case $\tilde\alpha =- \pi \frac{2D}{C}$. From (\ref{eqn:Srot-SL2}), we obtain
\begin{equation} \label{micro_crt}
    S_{\text{rot}}(E) =\frac{2 \pi  \sqrt{cE}}{\sqrt{3} C} = \frac{S_0}{C} + O(1).
\end{equation}
Even this expression is subleading with respect to the BTZ black hole. However, is has one important feature: it is purely real. Thus, its contribution to the density of states does not oscillate. It follows that if we consider the microcanonical ensemble with sufficiently wide energy windows, this is the contribution that will survive while all others may cancel out. We will see that this is exactly what happens in the thermal partition function. We are thus ready to answer the question posed in the Introduction: are there systematically more even than odd spin states? Looking at Eq. \eqref{micro_BTZ}, we see that this is not the case at leading order: indeed, $\textrm{Tr}_E \left( e^{\pi i J} \right)$ is a highly oscillatory function. However, the important point is that there is also a subleading contribution, given by \eqref{micro_crt}, which does not oscillate. This is the reason why there is a split between odd and even spins in the canonical ensemble.

\subsection{Phase diagrams}

We are now in a position to construct the phase diagram for $\text{Tr} \big( e^{- \beta H + i \tilde{\alpha} J} \big)$. The number and nature of the phases depend quite prominently on $\alpha$. We will consider only a few particular physically motivated values of that parameter. The phase diagram for arbitrary $\tilde{\alpha} \in \mathbb{Q}$ can be constructed using the phase diagrams discussed in \cite{Dijkgraaf:2000fq,Maloney:2007ud} and setting the modular parameter $\tau = \tilde{\alpha}$. In this sense, the results presented here are not new; we are rather reinterpreting them in light of the more recent work of \cite{Harlow:2023hjb, Benjamin:2024kdg}.

\begin{figure}[t]
\centering
\includegraphics[width=.8\textwidth]{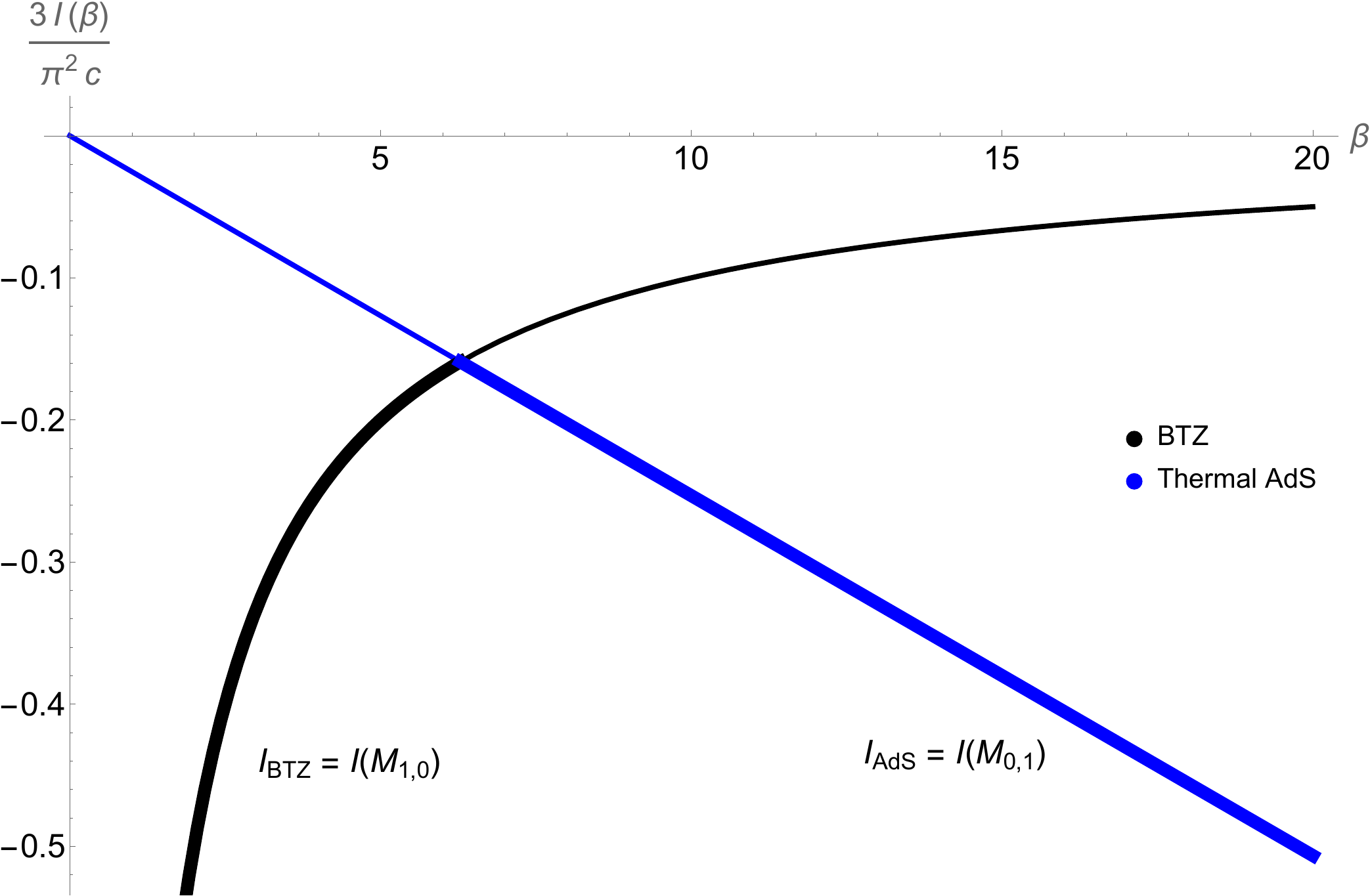}
\caption{The phase diagram for $\text{Tr}\left( e^{-\beta H} \right)$ in $\text{AdS}_3$ is shown. The competing saddles are thermal AdS (blue) and BTZ (black). The action $I(\beta)$ for both is plotted, with the dominant phase shown in bold. BTZ dominates at high temperatures, while thermal AdS dominates at low temperatures, with a Hawking--Page transition between them at $\beta_{\text{HP}} = 2\pi$.}
\label{fig:3d-thermal}
\end{figure}

Let us start with the well-known case $\tilde\alpha = 0$, i.e. a non-rotating system. We have two phases: the low-temperature behavior is governed by thermal $\text{AdS}_3$ ($M_{0,1}$), while at high temperatures the static BTZ black hole ($M_{1,0}$) dominates. The Hawking--Page transition between them occurs at $\beta = 2\pi$. This should not be a surprise, because these two saddles are connected by the $S$-transformation and the modular parameter $\tau = \frac{\tilde\alpha}{2\pi} + i \frac{\beta}{2\pi} = i$ is invariant under the $S$ transformations. This diagram is depicted in Figure~\ref{fig:3d-thermal}.

Now consider the case $\tilde\alpha = 2\pi$. By modular invariance, we should get the same phase diagram as for $\alpha = 0$. (However, this time the static BTZ black hole will correspond to the solution $M_{1,-1}$.) This is only true if the boundary CFT is purely bosonic. If it contains fermions, the modular group is smaller. In particular, for the Neveu-Schwarz (NS) sector, the only admissible saddles $M_{C,D}$ have $C+D$ is odd. This excludes $M_{1,-1}$, and can easily check that in this case $M_{0,1}$ (thermal AdS) dominates for all values of $\beta$. 

\begin{figure}[t]
\centering
\includegraphics[width=.8\textwidth]{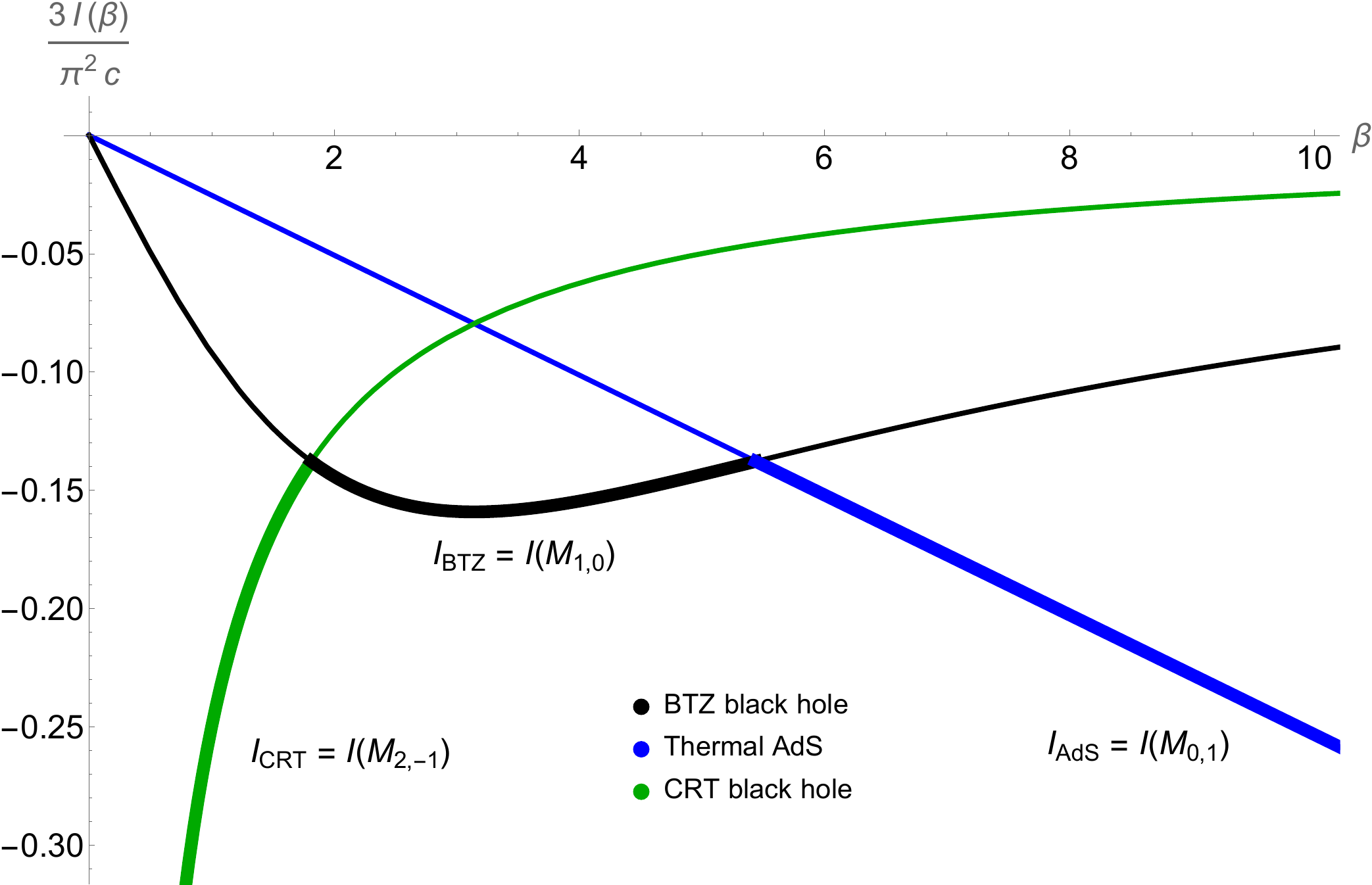}
\caption{The phase diagram for $\text{Tr}\left( e^{-\beta H + \pi i J} \right)$ in $\text{AdS}_3$ is shown. The competing saddles are thermal AdS (blue), rotating BTZ (black), and the $\mathcal{CRT}$-twisted black hole (green). The action $I(\beta)$ for all three is plotted, with the dominant phase shown in bold. The $\mathcal{CRT}$ solution dominates at high temperatures, BTZ at intermediate temperatures, and thermal AdS at low temperatures.}
\label{fig:3d-spinning}
\end{figure}

The value we are most interested in is $\tilde\alpha = \pi$. In this case, we will have three phases: thermal $\text{AdS}_3$, rotating BTZ ($M_{1,0}$ and $M_{1,-1}$) and the $\mathcal{CRT}$-twisted black hole $M_{2,-1}$. If there are NS fermions, the phase diagram is the same, except that $M_{1,-1}$ will drop out. The Hawking--Page transition now occurs at $\beta = \sqrt{3} \pi$, and the $\mathcal{CRT}$ saddle starts dominating at $\beta = \frac{\pi}{\sqrt{3}}$. These two temperatures are connected by the modular transformation 
\begin{equation}
    \tau \mapsto \frac{\tau - 1}{2 \tau -1}.
\end{equation}
Under that modular transformation, the low- and high-temperature regimes are exchanged via $\beta \mapsto \frac{\pi^2}{\beta}$. At high temperatures, we have
\begin{equation}
    \text{Tr} \left(e^{-\beta H + \pi i J} \right) = \exp\left(
\frac{c\pi^2}{12\beta^2}
    \right),
\end{equation}
which is very large in the semiclassical $(c \to \infty)$ limit. See Figure~\ref{fig:3d-spinning} for the phase diagram.

\begin{figure}[t]
\centering
\includegraphics[width=.8\textwidth]{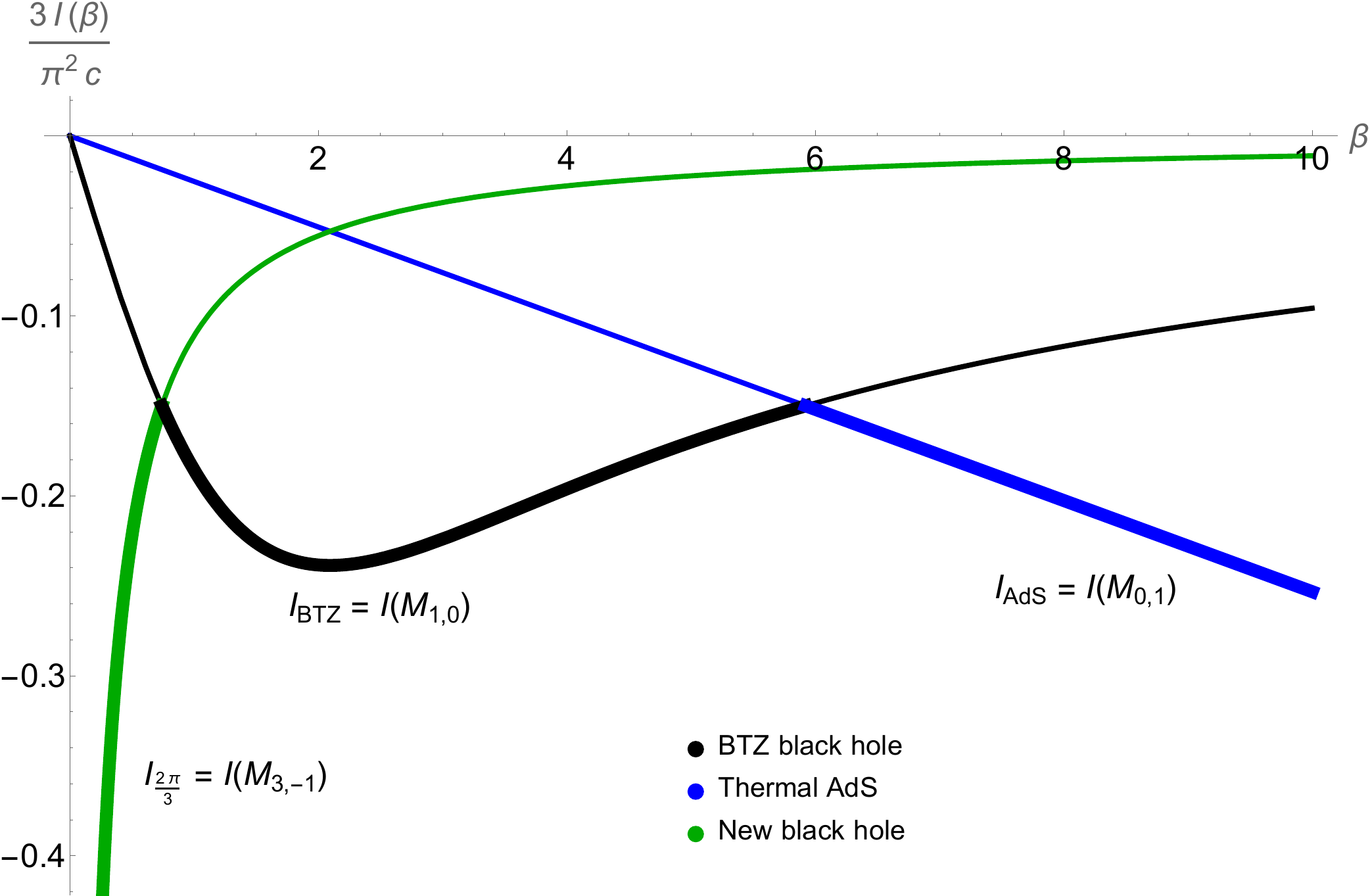}
\caption{The phase diagram for $\text{Tr}\left( e^{-\beta H + 2\pi i J/3} \right)$ in $\text{AdS}_3$ is shown. The competing saddles are thermal AdS (blue), rotating BTZ (black), and the $SL(2,\mathbb{Z})$ black hole $M_{3,-1}$ (green). The action $I(\beta)$ for all three is plotted, with the dominant phase shown in bold. The $SL(2,\mathbb{Z})$ solution dominates at high temperatures, BTZ at intermediate temperatures, and thermal AdS at low temperatures.}
\label{fig:3d-new}
\end{figure}

As a last example, let us consider $\tilde\alpha = \pm \frac{2\pi}{3}$. We have three phases: thermal $\text{AdS}_3$ ($M_{0,1}$), rotating BTZ ($M_{1,0}$), and the non-trivial $SL(2,\mathbb{Z})$ solution $M_{3,-1}$ (or $M_{3,1}$ for $\tilde\alpha = -\frac{2\pi}{3}$). The phase diagram for this case is shown in Figure~\ref{fig:3d-new}. At large temperatures, we find
\begin{equation}
    \text{Tr} \left( e^{-\beta H + i \frac{2\pi}{3} J} \right) = \exp\left(
\frac{c\pi^2}{27\beta^2}
    \right),
\end{equation}
which again goes to infinity as $\beta \to 0$.
The two values of the parameters $\tilde\alpha = \frac{2\pi}{3}$ and $\tilde\alpha = -\frac{2\pi}{3}$ are mapped into each other by the modular transformation
\begin{equation}
    \tau \mapsto \frac{-\tau}{3 \tau -1},
\end{equation}
which also takes $\beta \mapsto \frac{4\pi^2}{9 \beta}$. However, this modular transformation is not allowed in the presence of NS fermions, and neither is $M_{3,-1}$. In this case, at large temperatures the BTZ black hole still dominates. In particular, partition function remains bounded as $\beta \to 0$.

Let us now consider the case of generic $\tilde \alpha \in 2 \pi \mathbb{Q}$. We will not describe the full phase diagram, only its large temperature behavior. Let us write $\tilde\alpha = 2 \pi n/m$ with $m,n$ coprime and $m > 0$. To determine the phase structure, we want to minimize the function
\begin{equation}
    F_{\text{rot}} = -\frac{\pi^2 c}{3} \frac{m^2}{4\pi^2(n C + mD)^2 + m^2 C^2 \beta^2}
\end{equation}
at small $\beta$ over integers $C,D$. This can clearly be achieved if we choose $(C,D) = (m,-n)$. Then, we get
\begin{equation}
    F_{\text{rot}} = -\frac{\pi^2 c}{3} \frac{1}{m^2 \beta^2}.
\end{equation}
This assumes arbitrarily negative values as $\beta \to 0$. There is, however, a catch if we include fermions. Then, the saddle $M_{m,-n}$ does not satisfy the boundary conditions if $m -n$ is even. (Since they are coprime, it follows that both must be odd.) In this case, $F_{\text{rot}}$ must remain bounded as we go to $\beta \to 0$. In particular, in the limit it reads
\begin{equation}
    F_{\text{rot}} = -\frac{\pi^2 c}{3} \frac{m^2}{4\pi^2(n C + mD)^2}.
\end{equation}
Since $m,n$ are coprime, there exist by Euclid's algorithm integers $C,D$ such that $n C + mD = \pm 1$. Moreover, at least one of them is even. Thus, at high temperatures we have
\begin{equation}
    F_{\text{rot}} = -\frac{\pi^2 c}{3} \frac{m^2}{4\pi^2 + m^2 C^2 \beta^2}.
\end{equation}
The pair $(C,D)$ is unique up to $(C,D) \mapsto (C+m, D-n)$. We choose such $(C,D)$ that $C^2$ is as small as possible and that choice gives us the dominating saddle. 

Let us finish by discussing the case $\alpha \notin 2 \pi \mathbb{Q}$. For simplicity, we will restrict ourselves to the bosonic case. In this case the high-temperature limit of $F_{\text{rot}}$ (associated to $M_{C,D}$) is
\begin{equation}
    F_{\text{rot}}(\beta \to 0) = - \frac{\pi^2 c}{3} \frac{1}{(\tilde \alpha C + 2\pi D)^2},
\end{equation}
and we cannot kill the denominator. However, since the irrational numbers can be approximated arbitrarily well by rational ones, we can make the denominator as close to zero as we want by an appropriate choice of $(C,D)$. Thus, there will be no single dominant saddle at high temperatures. Instead, we will probe infinitely many different phases as we change the temperature. Using Hurwitz's theorem, one can show that in fact
\begin{equation}
    F_{\text{rot}}(\beta) \sim -\frac{c}{\beta}.
\end{equation}

Indeed, Hurwitz's theorem states that we may approximate any irrational number $\xi$ by a rational one, written in lowest terms as $\frac{m}{n}$, in such a way that
\begin{equation}
    \left\vert \xi - \frac{m}{n}\right\vert < \frac{1}{\sqrt{5}n^2}.
\end{equation}
Now the free energy associated to a sadle $M_{C,D}$ is given by
\begin{equation}
   F_{\text{rot}}(\tilde\alpha,\beta) = - \frac{\pi^2 c}{3C^2} \times \frac{1}{\tilde\beta^2 + \left(
    \xi + \frac{D}{C}
    \right)^2},
\end{equation}
where $\tilde\beta = \frac{\beta}{2\pi}$ and $\xi = \frac{\tilde\alpha}{2\pi}$ is any irrational number. By Hurwitz's theorem, we may choose $C,D$ so that
\begin{equation}
    \left( \xi + \frac{D}{C} \right)^2 < \frac{1}{5 C^4}.
\end{equation}
It follows that for this choice of $C,D$ we have
\begin{equation}
    F_{\text{rot}}(\tilde\alpha,\beta) < - \frac{\pi^2 c}{3C^2} \times \frac{1}{\tilde\beta^2 +\frac{1}{5 C^4} }.
\end{equation}
As long as we take $C \sim \tilde\beta^{-1/2}$, this quantity is of order $-\frac{c}{\beta}$.

This implies that the partition function with this insertion scales like $e^{\# c}$ at high temperatures. For a similar discussion using continued fractions, see \cite{Benjamin:2024kdg}.

\subsection{Introducing fermions}
So far, we have restricted ourselves to bosonic theories. Let us now briefly discuss the inclusion of fermions. In their presence, we must specify also the choice of the spin structure on the boundary. Then, in the gravitational path integral we sum over only those geometries that admit spin structures compatible with what was prescribed on the boundary.

In $\text{AdS}_3$, the boundary is simply a torus $S^1 \times S^1$. The choice of the spin structure is equivalent to the choice of either periodic or antiperiodic boundary conditions on each circle. Thus, we have four different spin structures. In the bulk, we must identify which cycles are contractible and demand that fermions are antiperiodic over them. We have four possibilities:
\begin{itemize}
    \item \textbf{Fermions are antiperiodic over both cycles.} This corresponds to the usual thermal partition function in the Neveu-Schwarz (NS) sector. The contributing saddles $M_{C,D}$ must have $C+D$ odd.
    \item \textbf{Fermions are periodic on the time circle and antiperodic on the spatial one.} In this case, we are calculating the index in the NS sector. The contributing saddles $M_{C,D}$ must have $D$ odd. 
    \item \textbf{Fermions are antiperiodic over the time circle and perodic over the spatial one.} In this case, we are calculating the thermal partition function in the Ramond (R) sector. The saddles contributing $M_{C,D}$ saddles have $C$ even. 
    \item \textbf{Fermions are periodic over both cycles.} In this case, we are calculating the index in the R sector. There is no smooth saddle compatible with these conditions: at least one circle always shrinks to zero.
\end{itemize}
Note that the $\mathcal{CRT}$-twisted black hole is $M_{2,-1}$, and as such contributes to the first three partition functions. One may also check that the first three partition functions are all connected by simple $SL(2, \mathbb{Z})$ transformations. Indeed, we have

\begin{equation}
    \textrm{Tr}_\textrm{NS} \left( e^{-\beta H + i \tilde\alpha J} (-1)^{\mathsf{F}} \right) = \textrm{Tr}_\textrm{NS} \left( e^{-\beta H + i (\tilde\alpha + 2\pi) J} \right)
\end{equation}
and
\begin{equation}
    \textrm{Tr}_\textrm{R} \left( e^{-\beta H + i \tilde\alpha J} \right) = \textrm{Tr}_\textrm{NS} \left( e^{\frac{2\pi}{\tilde\alpha^2 + \beta^2} \left(-\beta H - i \tilde\alpha J \right)} (-1)^{\mathsf{F}} \right).
\end{equation}
The latter equation states that these two quantities are connected by an S-transformation.

%%%%%%%%%%%%%%%%%%%%%
\section{Rotating black holes in $\text{AdS}_5$} \label{5d}

\subsection{Which saddles could contribute?}

In this section, we study spin-refined partition functions in holographic 4d CFTs dual to $\text{AdS}_5$ gravity by Euclidean path integral methods. We will begin by identifying the saddles that contribute to $\text{Tr}\left(e^{-\beta H + \beta\Omega_1 J_1 + \beta\Omega_2 J_2} \right)$ for fixed $\beta\Omega_1$ and $\beta\Omega_2$. Eventually, we will restrict ourselves to a few concrete values of these parameters. Then we will compute the Euclidean action and entropy of each saddle, examine their dominance in the canonical and microcanonical ensembles, respectively, and use the results to describe the phase diagrams. The relevant saddles we consider fall into three classes: (1) thermal AdS, (2) rotating black holes with complex angular momentum, and (3) quotients of Schwarzschild-AdS black holes, including $\mathcal{CRT}$-twisted black holes.

\paragraph{Thermal AdS.} Thermal AdS is the most obvious saddle, and always contributes to any spin-refined partition function. We choose conventions where its free energy is set to zero: this is equivalent to shifting all energies by the constant Casimir term $3\pi^2 \ell^2/32$, where $\ell$ is the AdS radius. In the microcanonical ensemble, the entropy of thermal AdS vanishes at leading order in $1/G_{\text{N}}$. (A more careful treatment of the entropy of the thermal gas would give a nonzero result due to one-loop effects.)

\paragraph{Rotating black holes.} The Myers--Perry black holes in $\text{AdS}_5$ are characterized by their mass $m$ and two spin parameters $a,b$ that govern their rotation about two axes \cite{Chong:2005hr}. The metric is given in Boyer--Lindquist coordinates $(t,r,\theta,\phi,\psi)$ by\footnote{We follow the conventions of \cite{Chen:2023mbc}. The careful reader may notice a small typo therein, where the horizon's area was given as the entropy.}
\begin{equation}
\label{eqn:MP-metric}
\begin{aligned}
\dd s^2 = &-\f{\Delta_{\theta} \left(1 + \f{r^2}{\ell^2} \right)}{\Xi_a \Xi_b} \dd t^2 + \f{2m}{\rho^2} \left( \f{\Delta_{\theta}\, \dd t}{\Xi_a \Xi_b} - \f{a}{\Xi_a} \sin^2\theta\, \dd\phi - \f{b}{\Xi_b} \cos^2\theta\, \dd\psi \right)^2 \\ &+ \rho^2 \left( \f{\dd r^2}{\Delta_r} + \f{\dd \theta^2}{\Delta_{\theta}} \right) + \left(\f{r^2 + a^2}{\Xi_a}\right) \sin^2\theta\, \dd\phi^2 + \left(\f{r^2 + b^2}{\Xi_b}\right) \cos^2\theta\, \dd\psi^2,
\end{aligned}
\end{equation}
where we have defined $\rho^2 = r^2 + a^2 \cos^2\theta + b^2 \sin^2\theta$, $\Xi_a = 1 - \f{a^2}{\ell^2}$, $\Xi_b = 1 - \f{b^2}{\ell^2}$, and
\begin{align}
\label{eqn:Delta}
\Delta_r = \f{1}{r^2} \left(r^2 + a^2 \right) \left(r^2 + b^2 \right) \left(1 + \f{r^2}{\ell^2} \right) - 2m, \qquad
\Delta_{\theta} = 1 - \f{a^2}{\ell^2} \cos^2\theta - \f{b^2}{\ell^2} \sin^2\theta.
\end{align}
The horizon radius $r_+$ is defined as the largest root of $\Delta_r$. This allows us to write the mass as $m = \f{1}{2r_+} \left(r_+^2 + a^2\right) \left(r_+^2 + b^2\right) \big(1 + \f{r_+^2}{\ell^2}\big)$. Thus we can label distinct solutions by $(a, b, r_+)$. The energy and the angular momenta generating rotations in the $\phi$ and $\psi$ directions are
\begin{align}
\label{eqn:MP-charges}
E = \f{\pi m \left(2\Xi_a + 2\Xi_b - \Xi_a \Xi_b \right)}{4 \Xi_a^2 \Xi_b^2}, \qquad
J_1 = \f{\pi a m}{2 \Xi_a^2 \Xi_b}, \qquad
J_2 = \f{\pi b m}{2 \Xi_a \Xi_b^2}.
\end{align}
Finally, the inverse temperature and angular velocities conjugate to $J_1$ and $J_2$, obtained by demanding that the Euclidean solution is smooth at the horizon, are given by
\begin{align}
\label{eqn:MP-potentials}
\beta = \f{2\pi r_+ \left(r_+^2 + a^2\right)\left(r_+^2 + b^2\right)}{r_+^4 \left( 1 + \f{1}{\ell^2} \left(2r_+^2 + a^2 + b^2 \right)\right)}, \qquad
\Omega_1 = \f{a \left(1 + \f{r_+^2}{\ell^2} \right)}{r_+^2 + a^2}, \qquad
\Omega_2 = \f{b \left(1 + \f{r_+^2}{\ell^2} \right)}{r_+^2 + b^2}.
\end{align}

These black holes contribute to a gravitational path integral for $\text{Tr} \left( e^{-\beta H + \beta\Omega_1 J_1 + \beta \Omega_2 J_2} \right)$. By holding the potentials $\beta \Omega_1$ and $\beta \Omega_2$ fixed and purely imaginary, we can identify saddles that contribute to certain spin-refined densities of states. For each saddle, we will express the parameters $(a, b, r_+)$ as functions of $\beta$ or $E$, depending on the ensemble, and then use (\ref{eqn:MP-charges}) and (\ref{eqn:MP-potentials}) to compute their on-shell action and entropy:
\begin{align}
\label{eqn:MP-action-entropy}
I_{\text{BH}} = \beta E - \f{A}{4} - \beta \Omega_1 J_1 - \beta \Omega_2 J_2, \qquad S_{\text{BH}} = \f{A}{4}, \qquad A = \f{2\pi^2 \left(r_+^2 + a^2\right) \left(r_+^2 + b^2\right)}{r_+ \Xi_a \Xi_b}.
\end{align}

\paragraph{Quotients of Schwarzschild.} As discussed in the Introduction, one may construct a $\mathcal{CRT}$-twisted black hole in any dimension via a smooth $\mathbb{Z}_2$ quotient of the Schwarzschild AdS black hole. This solution always contributes to $Z_R = \text{Tr} \left( e^{-\beta H} R \right)$, with Euclidean action (\ref{I_CRT}) and entropy (\ref{eqn:CRT-entropy}). As we will see, this saddle is dominant for $Z_R$ at high temperatures. 

For generic rational angular potentials $\alpha = 2\pi i p/q$, a family of $\mathbb{Z}_q$ quotients of Kerr AdS was recently constructed in \cite{Benjamin:2024kdg}. It was argued there that these quotients should dominate the corresponding spin-refined partition functions at high temperatures, at least provided that they are smooth. The $\mathcal{CRT}$-twisted black hole will be sufficient for our purposes, and we will not consider more general saddles here.

\subsection{Which saddles don't contribute?}

It is a longstanding open problem to systematically determine whether a given saddle will contribute to a gravitational path integral. The issue is especially complicated when the solution contains complex parameters. We defer to future work a more detailed investigation of contour prescriptions for such path integrals---see \cite{Marolf:2022ybi} for a proposal in this direction. In the meantime we will restrict ourselves to certain rules of thumb for the choices to be made. The best known is, of course, the Kontsevich--Segal--Witten criterion \cite{Kontsevich:2021dmb, Witten:2021nzp}. However, it is rather complicated to check in practice. Below, we discuss two simpler alternatives that will help us exclude certain complex saddles.

\paragraph{Thermal dominance criterion.} Suppose that at a given temperature $\beta$, the spin-refined partition function $\text{Tr}(e^{-\beta H + i\tilde{\alpha} J})$ (where $\tilde{\alpha} = -i\alpha \in \mathbb{R}$) is dominated by a single saddle with Euclidean action $I_{\alpha}(\beta)$, while the ordinary thermal partition function $\text{Tr}(e^{-\beta H})$ has dominant saddle-point contribution $I_0(\beta)$. Then we have
\begin{align}
\label{eqn:thermal-dominance}
\left\lvert \text{Tr}\left( e^{-\beta H + i\tilde{\alpha} J} \right) \right\rvert \leq \text{Tr} \left( \left\lvert e^{-\beta H + i\tilde{\alpha} J} \right\rvert \right) = \text{Tr} \left( e^{-\beta H} \right).
\end{align}
In the saddle-point approximation, the right-hand side is given by $e^{-I_0(\beta)}$, whereas the left-hand side is given by $|e^{-I_\alpha(\beta)}|$.\footnote{If there are multiple contributing saddles with the same real part, we should add their contributions. This changes the answer by an $O(1)$ factor, subleading in the $1/G_{\text{N}}$ expansion, so we will ignore this issue.} It follows that $\Re\left(I_{\alpha}(\beta)\right) \geq I_0(\beta)$, meaning that any putative saddle of the spin-refined partition function must have larger real part of the action than the corresponding saddle of the thermal partition function. For example, since thermal AdS dominates the canonical ensemble below the Hawking--Page temperature in the non-rotating case, we get a bound on rotating saddles in this regime: no solution that contributes to a spin-refined partition function may dominate over thermal AdS below the Hawking--Page temperature. In simple cases, (\ref{eqn:thermal-dominance}) is the statement that differences between certain numbers of states cannot exceed the total number of states. More generally, replacing differences with phases cannot increase the density of states either.

\paragraph{Horizon criterion.} There is a small subtlety that needs to be taken into account while discussing rotating black holes. We will soon solve certain algebraic equations to obtain $r_+$, for example as a function of the inverse temperature $\beta$. It will follow automatically from those equations that $r_+$ is a root of $\Delta_r$, as it should be. However, for real geometries $r_+$ must be the \textit{largest} root, and the equations we will solve make no such guarantee. We will encounter below an example where indeed certain real saddles can be excluded by that argument. Of course, it makes sense to talk about the ordering of roots only for real geometries. For more general complex saddles, it is natural to require that $r_+$ should be the root of $\Delta_r$ with the largest real part. We leave it to future work to determine whether this can be derived from a more principled formulation of the gravitational path integral.\footnote{By applying a complex diffeomorphism to the solution, one could potentially change the locations of the horizons in the complex plane. But this cannot be done arbitrarily, since otherwise one could switch the inner and outer horizons of a real geometry, which would change the solution.}

\subsection{The non-rotating case}

As a warm-up, and to showcase how we decide which solutions contribute, we review the computation of the thermal partition function $Z = \text{Tr}(e^{-\beta H})$. The relevant saddles are thermal AdS and the Schwarzschild $\text{AdS}_5$ black hole, which is obtained from (\ref{eqn:MP-metric}) by fixing $\beta \Omega_1 = \beta \Omega_2 = 0$. This sets $a=b=0$, so that black hole solutions are characterized by $r_+$.

\begin{figure}[t]
\centering
\includegraphics[width=.8\textwidth]{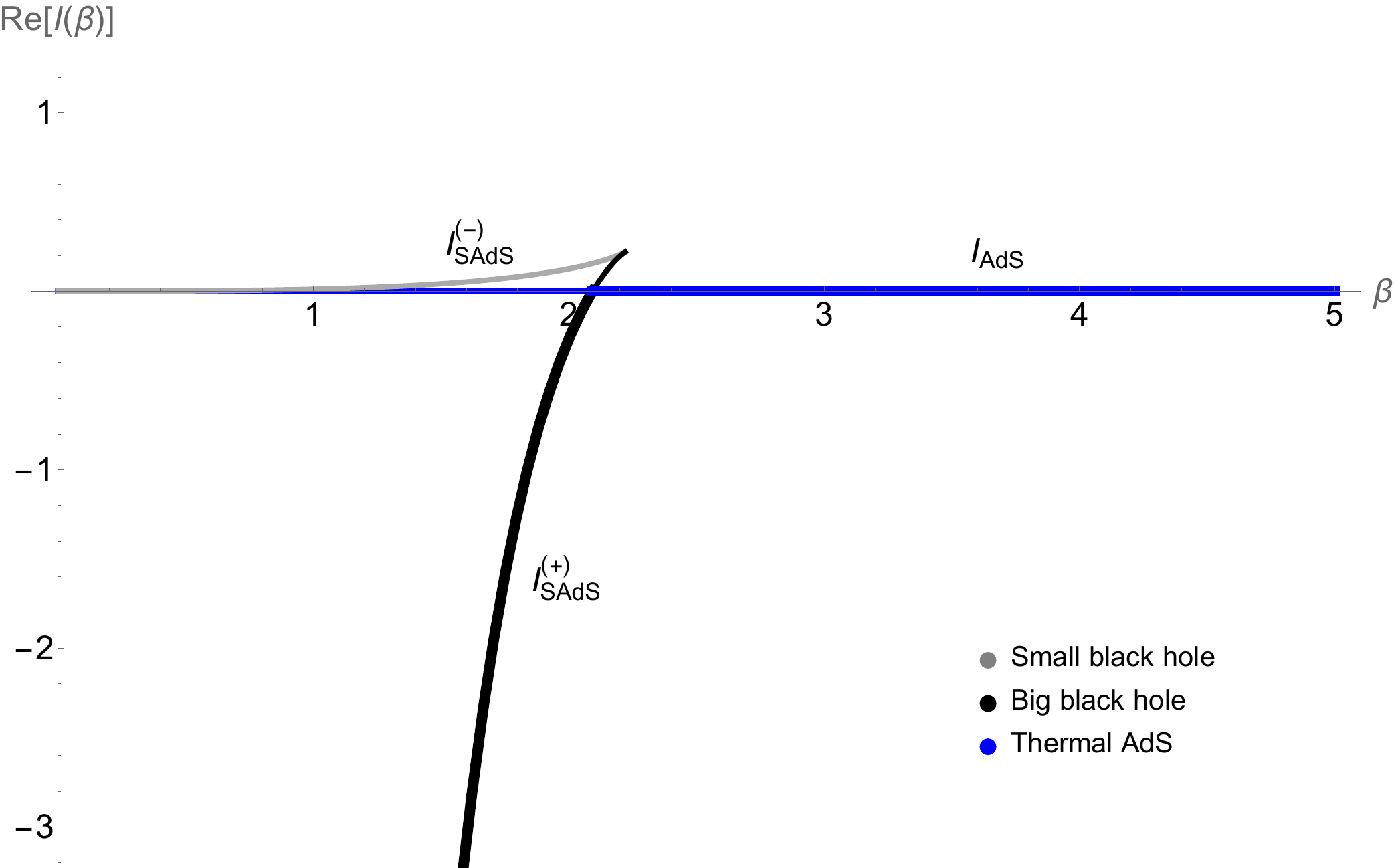}
\caption{The phase diagram for $\text{Tr}\left( e^{-\beta H} \right)$ in $\text{AdS}_5$ is shown. The relevant saddles are thermal AdS (blue) and two black hole solutions (gray and black). The action $\Re(I(\beta))$ for all three is plotted, with the dominant phase shown in bold. The big black hole dominates at high temperatures, while thermal AdS dominates at low temperatures, with a Hawking--Page transition between them at $\beta_{\text{HP}} = 2\pi \ell/3$. For this and all subsequent figures, we work in units where $\ell = 1$.}
\label{fig:thermal-canonical}
\end{figure}

In the canonical ensemble, by solving for $r_+$ in (\ref{eqn:MP-potentials}), one finds two solutions:
\begin{align}
\label{eqn:schwarz-rplus}
\beta = \f{2\pi \ell^2 r_+}{2r_+^2 + \ell^2} \implies
r_+(\beta) = \frac{\pi \ell^2}{2\beta} \left(1 \pm \sqrt{1 - \frac{2\beta^2}{\pi^2 \ell^2}} \right).
\end{align}
Both solutions define real geometries above a minimum temperature given by $\beta_{\text{min}} = \pi \ell/\sqrt{2}$, and it can be checked explicitly that both solutions satisfy the horizon criterion for all $\beta$. Their on-shell actions can be computed by plugging (\ref{eqn:schwarz-rplus}) into (\ref{eqn:MP-action-entropy}): one finds
\begin{align}
\label{eqn:SAdS-action}
I_{\text{SAdS}}^{(\pm)}(\beta) = -\f{\pi\ell^2}{64\beta} \left(\pi\ell \pm \sqrt{\pi^2\ell^2 - 2\beta^2}\right)^2 \left(\pi\ell \left(\pi\ell \pm \sqrt{\pi^2\ell^2 - 2\beta^2} \right) - 3\beta^2 \right).
\end{align}
This leads to the well-known phase diagram shown in Figure~\ref{fig:thermal-canonical}. The big black hole is the dominant saddle at high temperatures, where it has radius $r_+^{(+)}(\beta) \sim \pi\ell^2/\beta$ and action $I_{\text{SAdS}}^{(+)}(\beta) \sim -\pi^5 \ell^6/(8\beta^3)$. It remains dominant down to the Hawking--Page temperature $\beta_{\text{HP}} = 2\pi \ell/3$; below this temperature, thermal AdS dominates. The small black hole is subdominant for all temperatures where it exists.\footnote{Actually, at one-loop order one discovers that the small black hole has a negative mode \cite{Prestidge:1999uq,Marolf:2022jra}. This means that it cannot contribute to the gravitational path integral in the saddle point approximation.}

In the microcanonical ensemble, there are four solutions $r_+(E)$. From (\ref{eqn:MP-charges}), we find
\begin{align}
E = \f{3\pi}{8} r_+^2 \left(1 + \f{r_+^2}{\ell^2} \right) \implies
r_+^{(\pm, \pm)} = \pm \sqrt{\f{-\ell^2 \sqrt{3\pi} \pm \ell \sqrt{3\pi \ell^2 + 32E}}{2 \sqrt{3\pi}}}.
\end{align}
Two solutions are purely imaginary, and two are real. One of the latter has $r_+ < 0$, so it fails the horizon criterion and is excluded. Since from (\ref{eqn:MP-action-entropy}) we have $S = \pi^2 r_+^3/2$, the imaginary solutions have no real part of the entropy. This leaves only the real and positive solution: it exists for all energies, satisfies the horizon criterion, and dominates over thermal AdS at all energies. Its entropy grows at high energies like
\begin{align}
\label{eqn:SAdS-entropy}
S_{\text{SAdS}}(E) = \f{\pi^{5/4}}{12 \sqrt{6}} \left( \ell \sqrt{9\pi\ell^2 + 96E} - 3\ell^2\sqrt{\pi} \right)^{3/2} \sim \f{(2\pi)^{5/4} \ell^{3/2}}{3^{3/4}} E^{3/4} + O(E^{1/4}).
\end{align}

\subsection{Bosons minus fermions}

Let us consider the partition function with a $(-1)^{\mathsf{F}}$ insertion, which computes the difference between the number of bosonic and fermionic states in the CFT. This insertion was studied also in \cite{Chen:2023mbc}, and here we make only a few comments on their results. Due to the spin-statistic theorem, $(-1)^{\mathsf{F}} = e^{2 \pi i J}$, where $J$ is the angular momentum along any axis. Thus, the saddles we need to consider include thermal AdS and rotating black holes; for the latter, the $(-1)^{\mathsf{F}}$ insertion can be implemented either by $e^{2\pi i J_1}$ or $e^{2\pi i J_2}$, both of which count bosons and fermions with opposite signs. The answer does not depend on the choice, of course. For concreteness, we set $\beta\Omega_1 = 2\pi i$ and $\beta\Omega_2 = 0$, the latter of which fixes $b=0$. Black hole solutions will therefore be parametrized by pairs $(a, r_+)$.

The procedure for obtaining such solutions is straightforward. Using (\ref{eqn:MP-potentials}), the condition $\beta \Omega_1 = 2\pi i$ can be used to solve for $a$ in terms of $r_+$. One then substitutes this into the expressions (\ref{eqn:MP-charges}--\ref{eqn:MP-potentials}) for the inverse temperature or the energy, depending on the ensemble, and inverts to find $r_+(\beta)$ or $r_+(E)$ (and therefore also $a(\beta)$ or $a(E)$), respectively.

\begin{figure}[t]
\centering
\includegraphics[width=.8\textwidth]{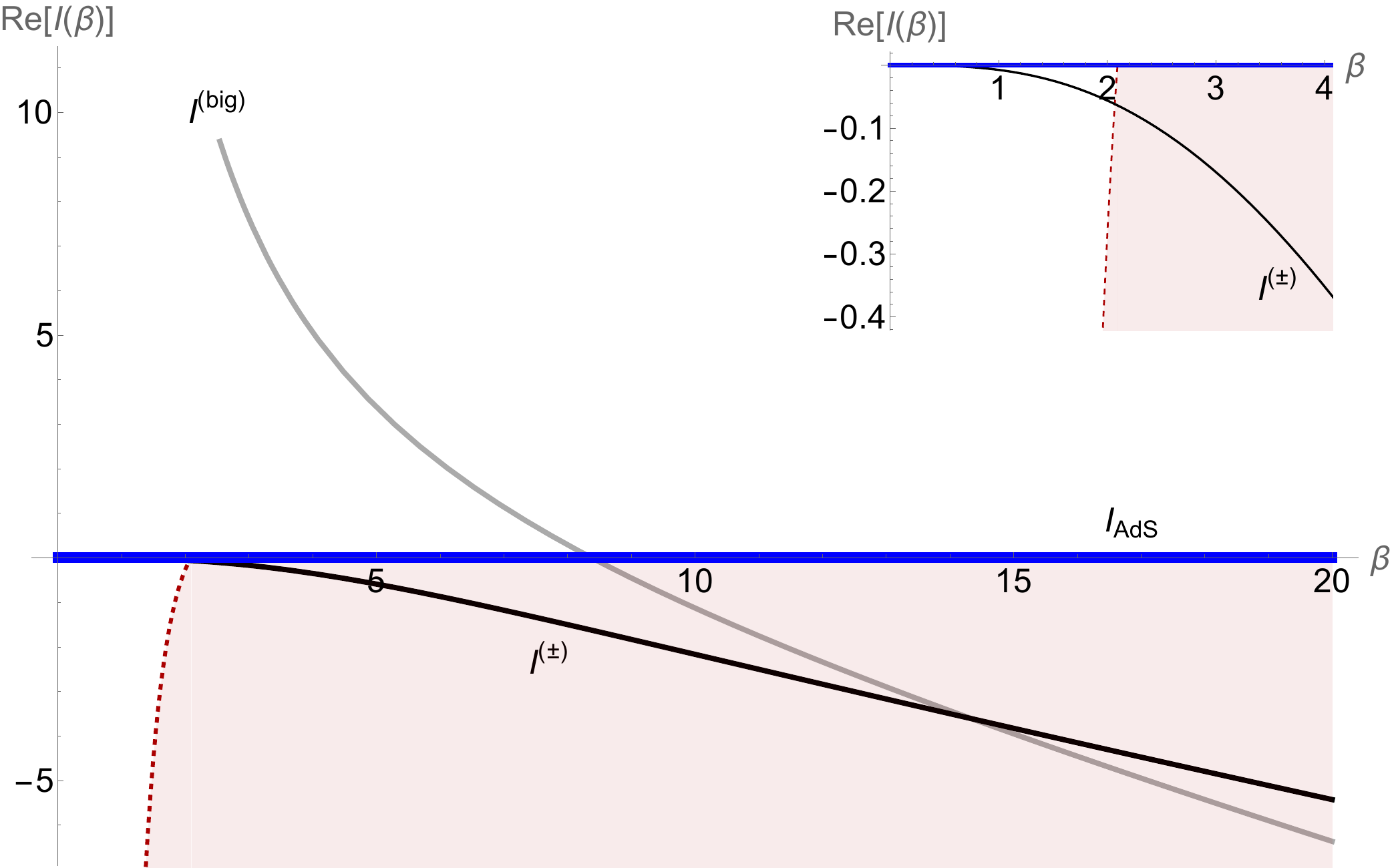}
\caption{The phase diagram for $\text{Tr} \left( e^{-\beta H} (-1)^{\mathsf{F}} \right)$ in $\text{AdS}_5$ is shown. The relevant saddles are thermal AdS (blue) and three rotating black holes (gray, black). The action $\Re(I(\beta))$ for all four is plotted, with the dominant phase in bold. We have superimposed the phase diagram of the thermal partition function (dashed red), and by the dominance criterion (\ref{eqn:thermal-dominance}) any saddle that contributes to a spin-refined partition function cannot occupy the ``thermal exclusion zone'' (shaded). The inset shows the high-temperature behavior of a pair of complex saddles, which na\"ively appear to dominate at small $\beta$. However, we expect them not to contribute, so the ensemble is always dominated by thermal AdS.}
\label{fig:index-canonical}
\end{figure}

This procedure can be carried out analytically in both the canonical and microcanonical ensembles. In the canonical ensemble, we get three explicit solutions $r_+(\beta)$, one real ``big'' black hole and two complex conjugates. At high temperatures, the real solution becomes large, while the complex solutions have vanishing real parts:
\begin{align}
r_+^{(\text{big})}(\beta) \sim \f{3\pi \ell^2}{\beta} + O(\beta), \qquad
r_+^{(\pm)}(\beta) \sim \pm \f{i\ell}{\sqrt{3}} + \f{\beta}{36\pi} + O(\beta^2).
\end{align}
We evaluate the real part of the on-shell action (\ref{eqn:MP-action-entropy}) explicitly for each of these solutions. The results are shown in Figure~\ref{fig:index-canonical}, which na\"ively indicates that these black holes could dominate, in disagreement with the conclusions of \cite{Chen:2023mbc}. At high temperatures, we have
\begin{align}
I^{(\text{big})}(\beta) \sim \f{27\pi^3}{32\beta} + O(\beta), \qquad
I^{(\pm)}(\beta) \sim \pm \f{i\beta^2}{12 \sqrt{3}} - \f{7\beta^3}{288\pi} + O(\beta^4).
\end{align}

However, none of these saddles can contribute to the gravitational path integral for two independent reasons. First, all three solutions fail the horizon criterion. Indeed, from (\ref{eqn:Delta}) the roots of $\Delta_r$ are $\pm r_+$ and $\pm \left(-a^2 - \ell^2 - r_+^2\right)^{1/2} =: \pm r_-$, and by plugging in the solutions $(a(\beta), r_+(\beta))$ one can show that for all three solutions, $\Re(r_-) > \Re(r_+)$ at all temperatures. In fact, for the big black hole, the metric and the roots of $\Delta_r$ are all real (after Wick rotation to Euclidean signature), so ``$r_+$'' is a genuine inner horizon. Second, as illustrated in Figure~\ref{fig:index-canonical}, the thermal dominance criterion requires legitimate saddles to lie outside of the shaded ``thermal exclusion zone.'' This prevents the big black hole from contributing at temperatures where it would dominate over thermal AdS, and the same criterion excludes the complex saddles at low temperatures. At high temperatures, these saddles cannot be excluded by this argument and appear to dominate the ensemble: see the inset in Figure~\ref{fig:index-canonical}. However, based on the horizon criterion and another argument that we give below, we conclude that they should not contribute to the gravitational path integral. It follows that thermal AdS gives the only allowed contribution to $\text{Tr}\left( e^{-\beta H + 2\pi i J} \right)$, and therefore dominates at all temperatures.

In the microcanonical ensemble, we get four solutions $(a(E), r_+(E))$. It can be shown that all four solutions have purely imaginary $r_+$, and therefore the entropy computed from (\ref{eqn:MP-action-entropy}) is also purely imaginary. To see this, note that $\beta\Omega_1 = 2\pi i$ has solution $a = -i\big( \f{\ell^2}{r_+} + 2r_+ \big)$, and substituting this into the condition $E = E(a, r_+)$ implies that $r_+$ must be a root of
\begin{equation}
    \frac{8E}{\pi} \left(\ell^2 + 4 r_+^2\right)^2 +(\ell^2+3r_+^2) \left(\ell^4 + 7 \ell^2 r_+^2 + 4 r_+^4 \right).
\end{equation}
Let us substitute $r_+^2 = -x$. The resulting cubic polynomial has three positive roots for $E>0$, as can be seen by checking its sign at $x=0, x=\frac{\ell^2}{4}, x=\frac{\ell^2}{3}$, and at infinity. From $a = -i\left( \f{\ell^2}{r_+} + 2r_+ \right)$ it follows that $a$ is purely real, so from (\ref{eqn:MP-action-entropy}) the entropy has no real part. The same conclusion was reached in \cite{Chen:2023mbc} numerically. Thus thermal AdS dominates the microcanonical ensemble, and we see that the spin-refined density of states is not exponentially large in $1/G_{\text{N}}$ at any energy. From this analysis, it is also clear that no black hole saddle can dominate in the canonical ensemble, as this would define a regime of energies where the spin-refined density of states is exponentially large.

\subsection{Even minus odd spin}

\begin{figure}[t]
\centering
\includegraphics[width=.8\textwidth]{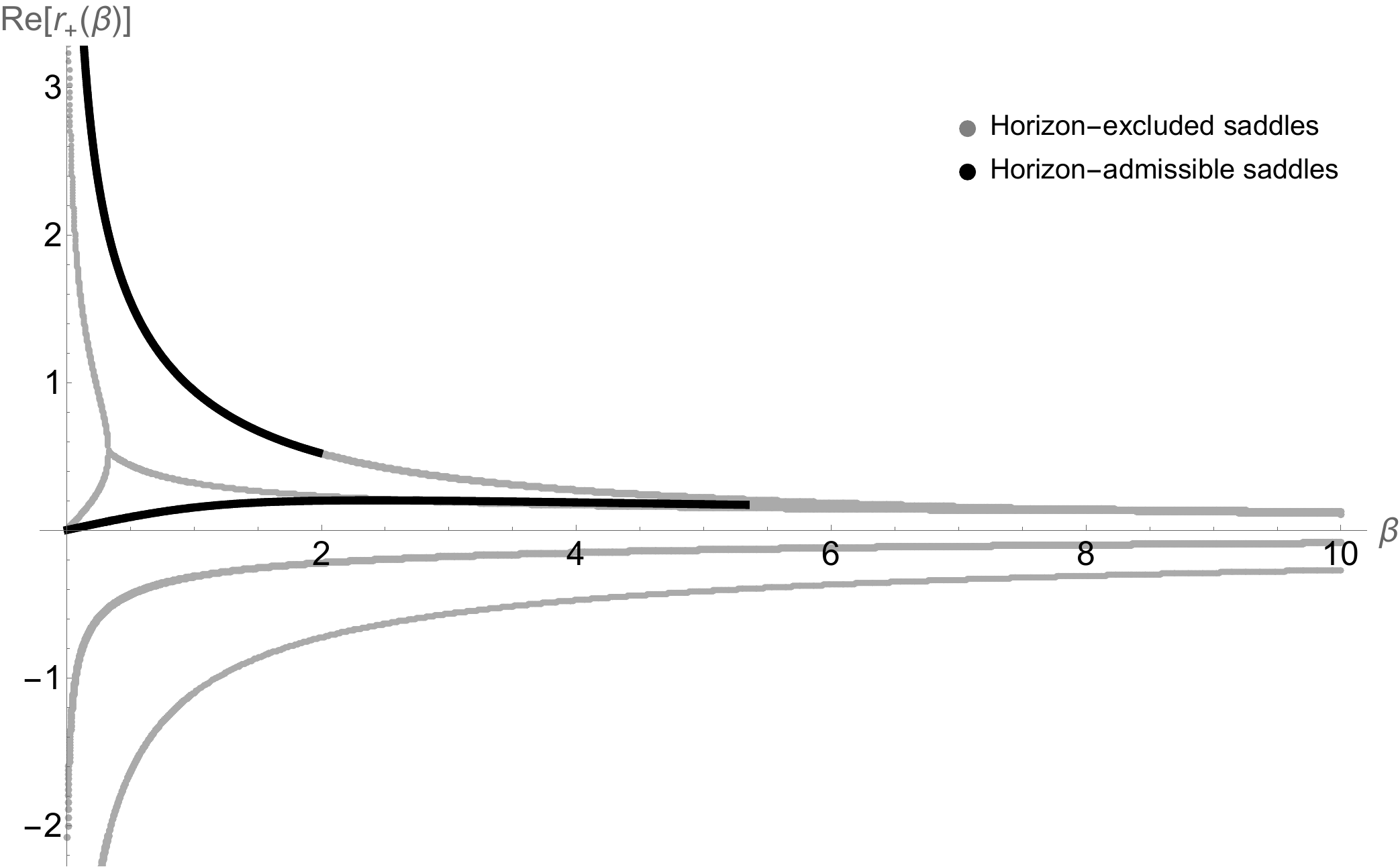}
\caption{At fixed temperature, there are 10 distinct rotating black holes in $\text{AdS}_5$ that satisfy $\beta\Omega_1 = \beta\Omega_2 = \pi i$. For each solution, we plot $\Re(r_+(\beta))$ and examine whether it is largest among the other roots of $\Delta_r$. Solutions are either excluded by this criterion (gray) or admissible (black).}
\label{fig:MP-canonical-horizon}
\end{figure}

The final example we consider is the partition function with an insertion of the reflection operator, $Z_R = \text{Tr} \left( e^{-\beta H} R \right)$. As described above, the relevant saddles are thermal AdS, rotating black holes, and now also the $\mathcal{CRT}$-twisted black hole. For the rotating solutions, the reflection operator is implemented by the rotation $e^{\pi i J_1 + \pi i J_2}$, so bulk solutions with $\beta\Omega_1 = \beta\Omega_2 = \pi i$ will contribute to $Z_R$. It follows that $a=b$, so the black hole solutions are again parametrized by $(a, r_+)$. The equations that determine them as functions of $\beta$ or $E$ define high-order polynomials, so we will have to proceed numerically.

Let us begin with the canonical ensemble. We found numerically that the system $\{\beta(a,r_+)\Omega_1(a,r_+) = \pi i,\; \beta = \beta(a,r_+)\}$ has 10 distinct solutions $(a(\beta), r_+(\beta))$. All of them come in complex conjugate pairs, except for one pair that bifurcates at high temperatures into two solutions with purely real $r_+$. The horizon criterion is only satisfied by two pairs of solutions, and even only above a minimal temperature for each one: see Figure~\ref{fig:MP-canonical-horizon}.

\begin{figure}[t]
\centering
\includegraphics[width=.8\textwidth]{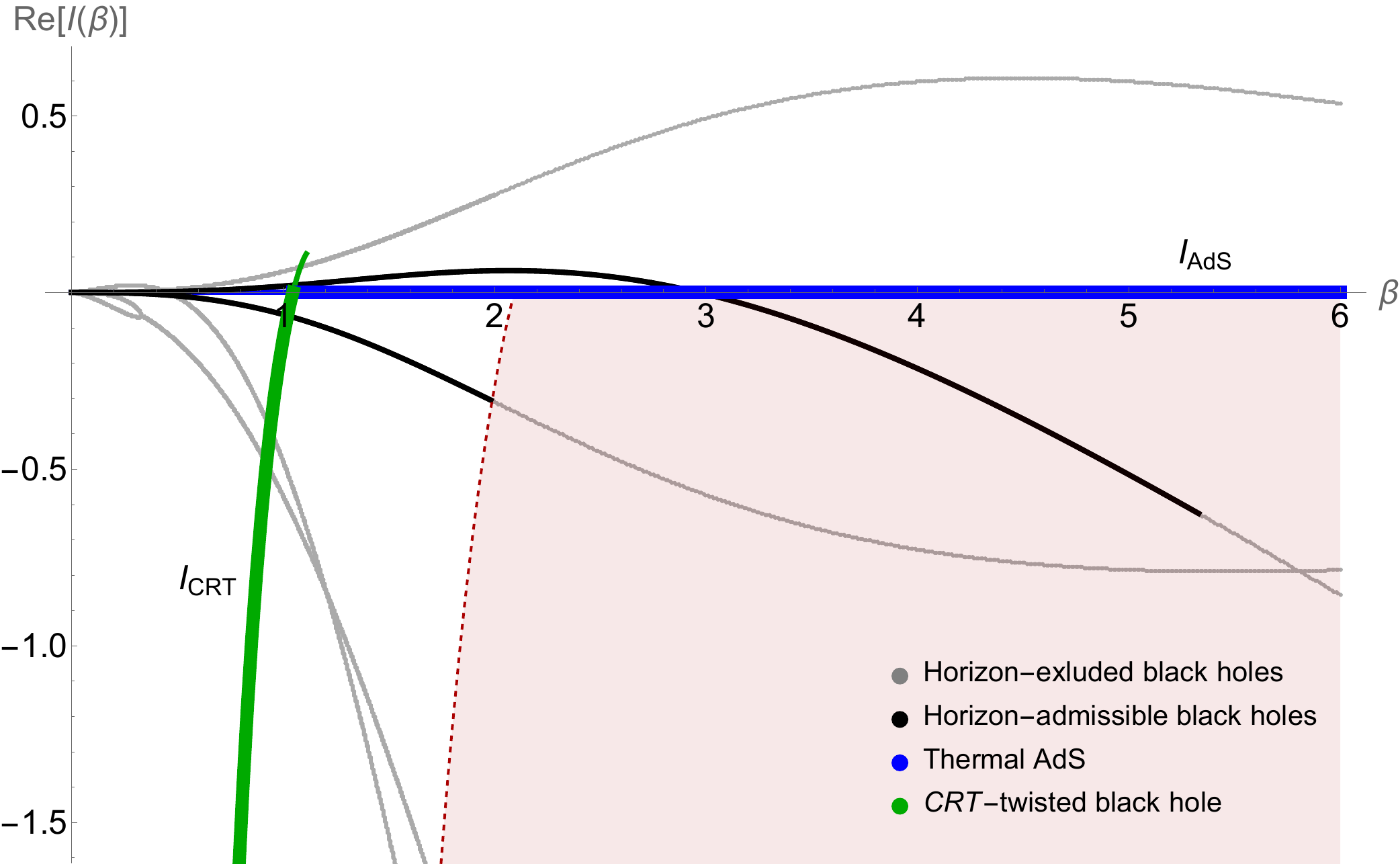}
\caption{The phase diagram for $\text{Tr} \left( e^{-\beta H} R \right)$ in $\text{AdS}_5$ is shown. The relevant saddles are thermal AdS (blue), 10 rotating black holes (gray or black, depending on the horizon criterion), and the $\mathcal{CRT}$ black hole (green). The action $\Re(I(\beta))$ for all of them is plotted, with the dominant phase in bold. Thermal AdS dominates at low temperatures, and the $\mathcal{CRT}$-twisted black hole dominates at high temperatures. The status of intermediate $\beta$ is complicated by the presence of a pair of complex saddles which are not immediately ruled out by our criteria. We nevertheless expect them not to contribute.}

\label{fig:MP-canonical-action}
\end{figure}

We numerically computed the Euclidean action of each of these solutions, which compete for dominance with thermal AdS and the $\mathcal{CRT}$-twisted black hole. The action of the latter is $I_{\mathcal{CRT}}(\beta) = \f{1}{2} I_{\text{SAdS}}^{(+)}(2\beta)$, with $I_{\text{SAdS}}^{(+)}(\beta)$ given by (\ref{eqn:SAdS-action}). The results are shown in Figure~\ref{fig:MP-canonical-action}, which also indicates the saddles that satisfy the horizon criterion and overlays the thermal exclusion zone below which no solution can contribute. Because thermal AdS always contributes, it must be the dominant phase above the Hawking--Page temperature $\beta_{\text{HP}} = 2\pi \ell/3$ by the thermal dominance criterion. At high temperatures, the $\mathcal{CRT}$-twisted black hole dominates and has $I_{\mathcal{CRT}}(\beta) \sim -\pi^5 \ell^6/(128 \beta^3)$ in this regime. It becomes subdominant to thermal AdS at $\f{1}{2}\beta_{\text{HP}} = \pi\ell/3$, and if no other saddles contribute at intermediate temperatures, then this new $\mathcal{CRT}$-twisted Hawking--Page transition fully characterizes the phase diagram of $Z_R$.

There is, in fact, one conjugate pair of black holes that dominate over both thermal AdS and the $\mathcal{CRT}$-twisted black hole and are not excluded from contributing by any of our criteria at intermediate temperatures.\footnote{Notice that both of our criteria begin to exclude this saddle at approximately the same temperature. We take this as evidence that our horizon criterion is correct, at least in the present setting.} So it may be possible for this regime to be dominated by a black hole. We should nevertheless exclude it, since (assuming we have not missed any other saddles) this would lead to a discontinuous spin-refined density of states.

\begin{figure}[t]
\centering
\includegraphics[width=.8\textwidth]{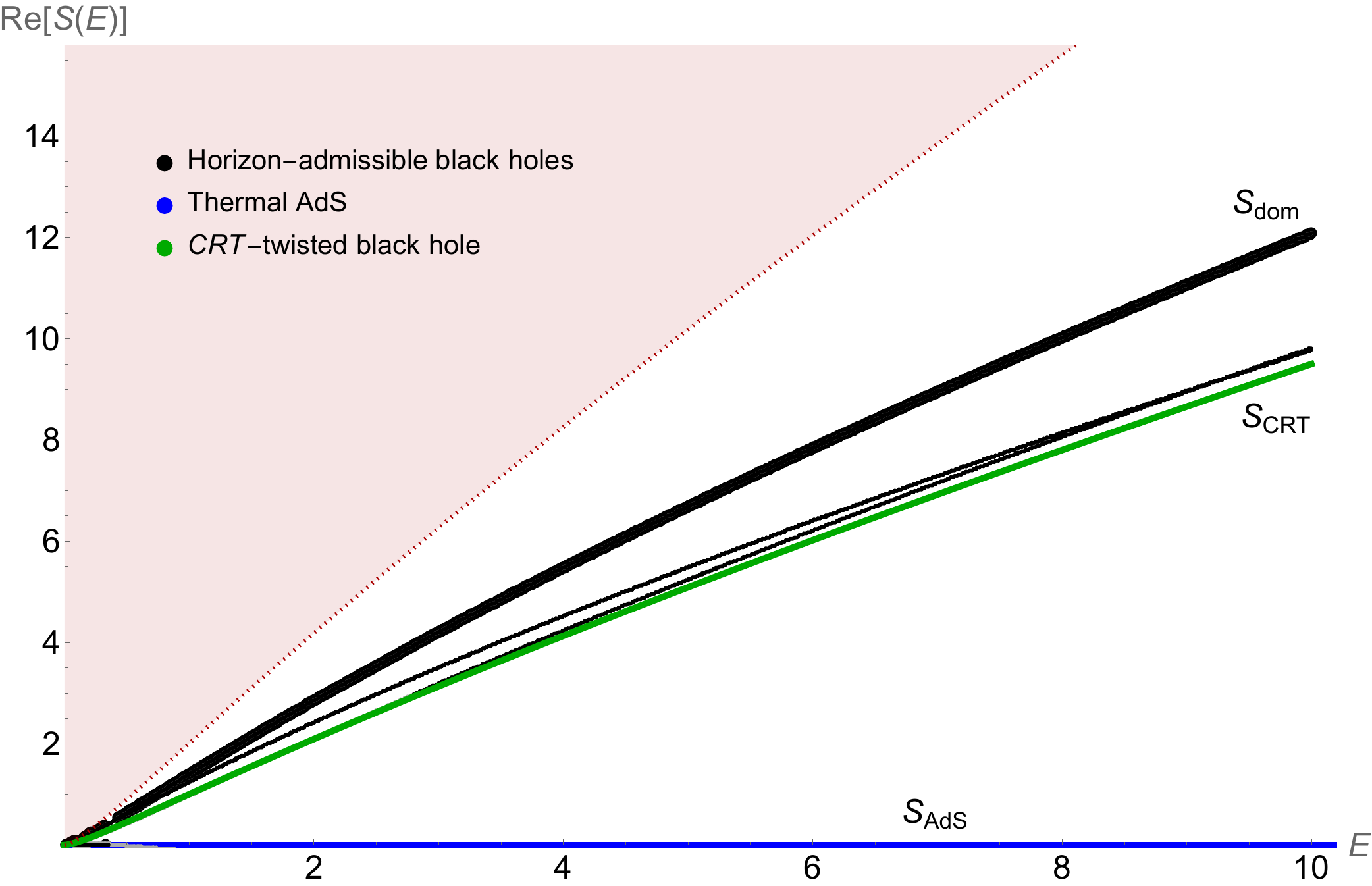}
\caption{The real part of the entropy counting the difference between even- and odd-spin states is shown as a function of the energy. Between thermal AdS (blue), the $\mathcal{CRT}$-twisted black hole (green), and a number of complex rotating black holes (black), one conjugate pair of black holes (bold) dominates at all energies. The exclusion zone (red) is also overlaid: valid contributions to $Z_R$ must have less entropy than the Schwarzschild black hole that dominates in the non-rotating case. We have checked that at larger energies the plot looks qualitatively the same.}
\label{fig:MP-microcanonical-entropy}
\end{figure}

A similar analysis can be repeated for the microcanonical ensemble. We found 30 distinct solutions $(a(E), r_+(E))$, all of which exist at all energies, and numerically computed the entropy $S(E)$ for each one. The results are shown in Figure~\ref{fig:MP-microcanonical-entropy}, alongside the entropy of thermal AdS and the $\mathcal{CRT}$-twisted black hole. The latter is just $S_{\mathcal{CRT}}(E) = \f{1}{2} S_{\text{SAdS}}(E)$, where $S_{\text{SAdS}}(E)$ is given by (\ref{eqn:SAdS-entropy}). Of those black hole solutions not excluded by the horizon criterion, one pair of complex saddles dominates the ensemble at all energies. We determined numerically that at high energies, their entropy grows like $\Re(S(E)) \sim 2.19 E^{3/4}$. Just as in the 3d case, these dominating saddles also have a nonzero imaginary part of the entropy, leading to a highly oscillatory microcanonical answer for the entropy. Meanwhile, the contribution of $\cal{CRT}$ black holes, while subdominant, is purely real.

%%%%%%%%%%%%%%%%%%%%%
\section{Discussion and further directions} \label{sec:Discussion}
%%%%%%%%%%%%%%%%%%%%%

We have investigated the r\^ole of $\mathcal{CRT}$-twisted black holes in the gravitational path integral and its microstate interpretations. In agreement with \cite{Benjamin:2024kdg}, we find that they give the leading contribution to the spin-refined partition function $\textrm{Tr} \big( e^{-\beta H} R \big)$, where $R$ is a reflection, at large temperatures. Using gravitational techniques, we are able to probe this quantity even at finite temperature. We are also able look more directly into the density of states by working in the microcanonical ensemble. We will review our main results below and point out some of the questions they raise.

\subsection{The microcanonical ensemble}

We have seen that the microcanonical ensemble is not equivalent to the canonical ensemble.\footnote{We have, however, used standard thermodynamic relations---which rely on ensemble equivalence---to compute the entropy and action of each saddle if it dominated. This reproduces the correct answers for $\mathcal{CRT}$-twisted and complex rotating black holes, even when they are not dominant. So ensemble equivalence is assumed to hold saddle by saddle, but which saddle dominates depends on the ensemble.} At large energies, the former is dominated by complex, rotating black holes, whereas at high temperatures the latter is dominated by the $\mathcal{CRT}$-twisted black hole. The reason for the non-equivalence of the ensembles is the rapid oscillations in the difference of densities of states for even and odd spins. These oscillations average out to give subdominant contributions to the thermal partition function.

We have checked that this is the case for $\text{AdS}_3$ and $\text{AdS}_5$. However, it is most likely not true in even-dimensional spacetimes. The reason behind this is just that antipodal map on $S^{D-2}$ is a rotation only for odd $D$. Thus, for even $D$ rotating black holes would not contribute to $\textrm{Tr}_E \big( e^{-\beta H} R \big)$. If we entertain the possibility that there are no other saddles that we are not aware of, then the $\mathcal{CRT}$-twisted black hole (which contributes in any dimension) dominates both the canonical and microcanonical ensembles in even dimension. This is an important qualitative difference between even and odd dimensions, and it would be interesting to see it arise directly in the CFT, for example using the thermal EFT approach of \cite{Benjamin:2024kdg}.

One could be  tempted to interpret the fact that $\Re( S(E)) = 0$ for the $\text{AdS}_5$ black holes contributing to $\textrm{Tr}\big( e^{-\beta H} (-1)^{\mathsf{F}} \big)$ as a sign of some emergent supersymmetry (perhaps at large energies) in holographic 4d CFTs. However, it was argued in \cite{Chen:2023mbc} that most likely there are additional saddles with defects that have larger entropy. A low-energy example of that type of construction was presented in \cite{Arkani-Hamed:2007ryu}. We plan to address the question of whether these types of geometries could also contribute to the spin-refined geometries in the future.

\subsection{Phase diagrams}

Using AdS/CFT, it is possible to probe the phase diagrams for spin-refined partition functions even at finite temperature. This is most clearly achieved in $\text{AdS}_3$, where we have a full list of all contributing saddles and thus we can draw the phase diagram for any rotational insertion, as was done previously in \cite{Dijkgraaf:2000fq, Maloney:2007ud}. Perhaps surprisingly, for such insertions one finds richer structures than in the case of the thermal partition function. For $\textrm{Tr} \big( e^{-\beta H} R \big)$ in $\text{AdS}_3$ we have three phases. At high temperatures, the dominating saddle is the $\mathcal{CRT}$-twisted black hole. At low temperatures this quantity is dominated by thermal AdS. There is, however, an intermediate regime in which the dominating saddle is a rotating BTZ black hole.

Interestingly, the exchange of the dominance between the latter two occurs at a temperature above the Hawking--Page transition. The points of transition are fixed by modular invariance,\footnote{We have presented only the simplest example. For the general construction of the phase diagram, see \cite{Dijkgraaf:2000fq}.} and thus we expect this feature to hold universally, not just for holographic theories. It would be interesting to get more detailed insight into these different phases intrinsically in the CFT. Of course, it also remains to be seen whether we can learn more about them on the gravity side. It would be especially interesting to see how the lack of time-orientability of $\mathcal{CRT}$-twisted black holes manifests itself in the $n$-point functions on the boundary. Since these topological issues are present only behind the horizon, this line of investigation would necessarily lead us outside of the realm of the Euclidean gravity.

We have also found and studied many saddles that potentially contribute to $\textrm{Tr} \big( e^{-\beta H} R \big)$ in $\text{AdS}_5$. The construction of the phase diagram is more complicated here because it is far from obvious which solutions are legitimate saddles that contribute to the gravitational path integral. Based on Figure~\ref{fig:MP-canonical-action}, we expect that no rotating black hole shall ever dominate.\footnote{A less desirable alternative would be that the spin-refined density of states is discontinuous in $\beta$.} On one hand, this is also what happens in $\text{AdS}_4$, where the reflection $R$ cannot be reduced to a rotation. On the other hand, we cannot exclude the possibility that there are more exotic saddles that we may have missed. To resolve these ambiguities, it would be beneficial to probe what happens at modest temperatures in the CFT directly. We also hope to find the precise contour-prescription for these quantitites to prove that all of the problematic saddles we have discussed can be rejected from first principles. 

%%%%%%%%%%%%%%%%%%%%%
\begin{acknowledgments}

It is a pleasure to thank Daniel Harlow, Don Marolf, Gary Horowitz, Joaquin Turiaci, David Berenstein, Xiaoyi Liu, Zixia Wei, Sridip Pal and Paweł Lewulis for helpful discussions. D.G. was supported in part by the Department of Energy under grant DE-SC 0011702. M.K. was supported in part by NSF Grant PHY-2107939.

\end{acknowledgments}

%%%%%%%%%%%%%%%%%%%%%%%%%%%%%%%%%%%%%%%%%%

\bibliography{bibliography}{}

\providecommand{\href}[2]{#2}\begingroup\raggedright\begin{thebibliography}{10}

\bibitem{Gibbons:1976ue}
G.~W. Gibbons and S.~W. Hawking, ``{Action Integrals and Partition Functions in
  Quantum Gravity},'' \href{http://dx.doi.org/10.1103/PhysRevD.15.2752}{{\em
  Phys. Rev. D} {\bfseries 15} (1977) 2752--2756}.

\bibitem{Cabo-Bizet:2018ehj}
A.~Cabo-Bizet, D.~Cassani, D.~Martelli, and S.~Murthy, ``{Microscopic origin of
  the Bekenstein-Hawking entropy of supersymmetric AdS$_{5}$ black holes},''
  \href{http://dx.doi.org/10.1007/JHEP10(2019)062}{{\em JHEP} {\bfseries 10}
  (2019) 062}, \href{http://arxiv.org/abs/1810.11442}{{\ttfamily
  arXiv:1810.11442 [hep-th]}}.

\bibitem{Iliesiu:2022kny}
L.~V. Iliesiu, S.~Murthy, and G.~J. Turiaci, ``{Black hole microstate counting
  from the gravitational path integral},''
  \href{http://arxiv.org/abs/2209.13602}{{\ttfamily arXiv:2209.13602
  [hep-th]}}.

\bibitem{Iliesiu:2022onk}
L.~V. Iliesiu, S.~Murthy, and G.~J. Turiaci, ``{Revisiting the Logarithmic
  Corrections to the Black Hole Entropy},''
  \href{http://arxiv.org/abs/2209.13608}{{\ttfamily arXiv:2209.13608
  [hep-th]}}.

\bibitem{LopesCardoso:2022hvc}
G.~Lopes~Cardoso, A.~Kidambi, S.~Nampuri, V.~Reys, and M.~Rossell\'o, ``{The
  Gravitational Path Integral for $ N=4$ BPS Black Holes from Black Hole
  Microstate Counting},''
  \href{http://dx.doi.org/10.1007/s00023-023-01297-y}{{\em Annales Henri
  Poincare} {\bfseries 24} no.~10, (2023) 3305--3346},
  \href{http://arxiv.org/abs/2211.06873}{{\ttfamily arXiv:2211.06873
  [hep-th]}}.

\bibitem{H:2023qko}
A.~A. H., P.~V. Athira, C.~Chowdhury, and A.~Sen, ``{Logarithmic correction to
  BPS black hole entropy from supersymmetric index at finite temperature},''
  \href{http://dx.doi.org/10.1007/JHEP03(2024)095}{{\em JHEP} {\bfseries 03}
  (2024) 095}, \href{http://arxiv.org/abs/2306.07322}{{\ttfamily
  arXiv:2306.07322 [hep-th]}}.

\bibitem{Sen:2023dps}
A.~Sen, ``{Revisiting localization for BPS black hole entropy},''
  \href{http://arxiv.org/abs/2302.13490}{{\ttfamily arXiv:2302.13490
  [hep-th]}}.

\bibitem{Anupam:2023yns}
A.~H. Anupam, C.~Chowdhury, and A.~Sen, ``{Revisiting logarithmic correction to
  five dimensional BPS black hole entropy},''
  \href{http://dx.doi.org/10.1007/JHEP05(2024)070}{{\em JHEP} {\bfseries 05}
  (2024) 070}, \href{http://arxiv.org/abs/2308.00038}{{\ttfamily
  arXiv:2308.00038 [hep-th]}}.

\bibitem{Maldacena:1997re}
J.~M. Maldacena, ``{The Large N limit of superconformal field theories and
  supergravity},'' \href{http://dx.doi.org/10.4310/ATMP.1998.v2.n2.a1}{{\em
  Adv. Theor. Math. Phys.} {\bfseries 2} (1998) 231--252},
  \href{http://arxiv.org/abs/hep-th/9711200}{{\ttfamily arXiv:hep-th/9711200}}.

\bibitem{Witten:1998zw}
E.~Witten, ``{Anti-de Sitter space, thermal phase transition, and confinement
  in gauge theories},''
  \href{http://dx.doi.org/10.4310/ATMP.1998.v2.n3.a3}{{\em Adv. Theor. Math.
  Phys.} {\bfseries 2} (1998) 505--532},
  \href{http://arxiv.org/abs/hep-th/9803131}{{\ttfamily arXiv:hep-th/9803131}}.

\bibitem{Witten:1982df}
E.~Witten, ``{Constraints on Supersymmetry Breaking},''
  \href{http://dx.doi.org/10.1016/0550-3213(82)90071-2}{{\em Nucl. Phys. B}
  {\bfseries 202} (1982) 253}.

\bibitem{Sen:2012hv}
A.~Sen, ``{BPS Spectrum, Indices and Wall Crossing in N=4 Supersymmetric
  Yang-Mills Theories},'' \href{http://dx.doi.org/10.1007/JHEP06(2012)164}{{\em
  JHEP} {\bfseries 06} (2012) 164},
  \href{http://arxiv.org/abs/1203.4889}{{\ttfamily arXiv:1203.4889 [hep-th]}}.

\bibitem{Benjamin:2019stq}
N.~Benjamin, H.~Ooguri, S.-H. Shao, and Y.~Wang, ``{Light-cone modular
  bootstrap and pure gravity},''
  \href{http://dx.doi.org/10.1103/PhysRevD.100.066029}{{\em Phys. Rev. D}
  {\bfseries 100} no.~6, (2019) 066029},
  \href{http://arxiv.org/abs/1906.04184}{{\ttfamily arXiv:1906.04184
  [hep-th]}}.

\bibitem{Pal:2020wwd}
S.~Pal and Z.~Sun, ``{High Energy Modular Bootstrap, Global Symmetries and
  Defects},'' \href{http://dx.doi.org/10.1007/JHEP08(2020)064}{{\em JHEP}
  {\bfseries 08} (2020) 064}, \href{http://arxiv.org/abs/2004.12557}{{\ttfamily
  arXiv:2004.12557 [hep-th]}}.

\bibitem{Harlow:2021trr}
D.~Harlow and H.~Ooguri, ``{A universal formula for the density of states in
  theories with finite-group symmetry},''
  \href{http://dx.doi.org/10.1088/1361-6382/ac5db2}{{\em Class. Quant. Grav.}
  {\bfseries 39} no.~13, (2022) 134003},
  \href{http://arxiv.org/abs/2109.03838}{{\ttfamily arXiv:2109.03838
  [hep-th]}}.

\bibitem{Harlow:2023hjb}
D.~Harlow and T.~Numasawa, ``{Gauging spacetime inversions in quantum
  gravity},'' \href{http://arxiv.org/abs/2311.09978}{{\ttfamily
  arXiv:2311.09978 [hep-th]}}.

\bibitem{Streater:1989vi}
R.~F. Streater and A.~S. Wightman, {\em {PCT, spin and statistics, and all
  that}}.
\newblock 1989.

\bibitem{Maloney:2016gsg}
A.~Maloney and S.~F. Ross, ``{Holography on Non-Orientable Surfaces},''
  \href{http://dx.doi.org/10.1088/0264-9381/33/18/185006}{{\em Class. Quant.
  Grav.} {\bfseries 33} no.~18, (2016) 185006},
  \href{http://arxiv.org/abs/1603.04426}{{\ttfamily arXiv:1603.04426
  [hep-th]}}.

\bibitem{Wei:2024zez}
Z.~Wei, ``{Holographic Dual of Crosscap Conformal Field Theory},''
  \href{http://arxiv.org/abs/2405.03755}{{\ttfamily arXiv:2405.03755
  [hep-th]}}.

\bibitem{Chen:2023mbc}
Y.~Chen and G.~J. Turiaci, ``{Spin-statistics for black hole microstates},''
  \href{http://dx.doi.org/10.1007/JHEP04(2024)135}{{\em JHEP} {\bfseries 04}
  (2024) 135}, \href{http://arxiv.org/abs/2309.03478}{{\ttfamily
  arXiv:2309.03478 [hep-th]}}.

\bibitem{Benjamin:2024kdg}
N.~Benjamin, J.~Lee, S.~Pal, D.~Simmons-Duffin, and Y.~Xu, ``{Angular fractals
  in thermal QFT},'' \href{http://arxiv.org/abs/2405.17562}{{\ttfamily
  arXiv:2405.17562 [hep-th]}}.

\bibitem{Cardy:1986ie}
J.~L. Cardy, ``{Operator Content of Two-Dimensional Conformally Invariant
  Theories},'' \href{http://dx.doi.org/10.1016/0550-3213(86)90552-3}{{\em Nucl.
  Phys. B} {\bfseries 270} (1986) 186--204}.

\bibitem{Maloney:2007ud}
A.~Maloney and E.~Witten, ``{Quantum Gravity Partition Functions in Three
  Dimensions},'' \href{http://dx.doi.org/10.1007/JHEP02(2010)029}{{\em JHEP}
  {\bfseries 02} (2010) 029}, \href{http://arxiv.org/abs/0712.0155}{{\ttfamily
  arXiv:0712.0155 [hep-th]}}.

\bibitem{Dijkgraaf:2000fq}
R.~Dijkgraaf, J.~M. Maldacena, G.~W. Moore, and E.~P. Verlinde, ``{A Black hole
  Farey tail},'' \href{http://arxiv.org/abs/hep-th/0005003}{{\ttfamily
  arXiv:hep-th/0005003}}.

\bibitem{Chong:2005hr}
Z.~W. Chong, M.~Cvetic, H.~Lu, and C.~N. Pope, ``{General non-extremal rotating
  black holes in minimal five-dimensional gauged supergravity},''
  \href{http://dx.doi.org/10.1103/PhysRevLett.95.161301}{{\em Phys. Rev. Lett.}
  {\bfseries 95} (2005) 161301},
  \href{http://arxiv.org/abs/hep-th/0506029}{{\ttfamily arXiv:hep-th/0506029}}.

\bibitem{Marolf:2022ybi}
D.~Marolf, ``{Gravitational thermodynamics without the conformal factor
  problem: partition functions and Euclidean saddles from Lorentzian path
  integrals},'' \href{http://dx.doi.org/10.1007/JHEP07(2022)108}{{\em JHEP}
  {\bfseries 07} (2022) 108}, \href{http://arxiv.org/abs/2203.07421}{{\ttfamily
  arXiv:2203.07421 [hep-th]}}.

\bibitem{Kontsevich:2021dmb}
M.~Kontsevich and G.~Segal, ``{Wick Rotation and the Positivity of Energy in
  Quantum Field Theory},'' \href{http://dx.doi.org/10.1093/qmath/haab027}{{\em
  Quart. J. Math. Oxford Ser.} {\bfseries 72} no.~1-2, (2021) 673--699},
  \href{http://arxiv.org/abs/2105.10161}{{\ttfamily arXiv:2105.10161
  [hep-th]}}.

\bibitem{Witten:2021nzp}
E.~Witten, ``{A Note On Complex Spacetime Metrics},''
  \href{http://arxiv.org/abs/2111.06514}{{\ttfamily arXiv:2111.06514
  [hep-th]}}.

\bibitem{Prestidge:1999uq}
T.~Prestidge, ``{Dynamic and thermodynamic stability and negative modes in
  Schwarzschild-anti-de Sitter},''
  \href{http://dx.doi.org/10.1103/PhysRevD.61.084002}{{\em Phys. Rev. D}
  {\bfseries 61} (2000) 084002},
  \href{http://arxiv.org/abs/hep-th/9907163}{{\ttfamily arXiv:hep-th/9907163}}.

\bibitem{Marolf:2022jra}
D.~Marolf and J.~E. Santos, ``{Stability of the microcanonical ensemble in
  Euclidean Quantum Gravity},''
  \href{http://dx.doi.org/10.1007/JHEP11(2022)046}{{\em JHEP} {\bfseries 11}
  (2022) 046}, \href{http://arxiv.org/abs/2202.12360}{{\ttfamily
  arXiv:2202.12360 [hep-th]}}.

\bibitem{Arkani-Hamed:2007ryu}
N.~Arkani-Hamed, S.~Dubovsky, A.~Nicolis, and G.~Villadoro, ``{Quantum Horizons
  of the Standard Model Landscape},''
  \href{http://dx.doi.org/10.1088/1126-6708/2007/06/078}{{\em JHEP} {\bfseries
  06} (2007) 078}, \href{http://arxiv.org/abs/hep-th/0703067}{{\ttfamily
  arXiv:hep-th/0703067}}.

\end{thebibliography}\endgroup
\bibliographystyle{utphys-modified}

\end{document}